\documentclass[10pt]{article}
\usepackage[ansinew]{inputenc}
\usepackage{bm}
\usepackage[dvips]{graphicx}
\usepackage{color}

\title{Geometrical locus of massive test particle orbits in the space of physical parameters in Kerr space-time.}
\author{F. Fayos\thanks{Also at Laboratori de F\'{\i}sica Matem\`atica,
Societat Catalana de F\'{\i}sica, IEC, Barcelona.} , Ch. Teij\'on \\
Department of Applied Physics, UPC, Barcelona, Spain. \\
\\}
\date{}

\begin{document}
\maketitle

\begin{abstract}
Gravitational radiation of binary systems can be studied by using
the adiabatic approximation in General Relativity. In this approach a small
astrophysical object follows a trajectory consisting of a chained series of bounded
geodesics (orbits) in the outer region of a Kerr Black Hole,
representing the space time created by a bigger object. In our paper
we study the entire class of orbits, both of constant radius (spherical orbits),
as well as non-null eccentricity orbits, showing a number of properties on the physical parameters and
trajectories.
The main result is the determination of the geometrical locus of all the orbits
in the space of physical parameters in Kerr space-time. This becomes a powerful tool to
know if different orbits can be connected by a continuous change of their physical parameters.
A discussion on the influence of different values of the angular
momentum of the hole is given. Main results have been obtained
by analytical methods.\\

\end{abstract}
\maketitle

\section{Introduction}

Orbits (geodesics with null or non null eccentricity) in the outer
region of Kerr space time play an important role in the description
of important astrophysical phenomena near black holes. This
space-time, characterized by the parameters $M$ and $a$ interpreted
as the mass and angular momentum per unit of mass of a rotating
hole, can be described in the outer region (limited by the
outer horizon and the asymptotically flat region) by the Boyer-Linquist coordinate $x^\alpha
\equiv \{ t, r, \theta, \phi \}$ .

Geodesics are characterized by four constants of motion, that are $E
, m, L$ and $\mathcal{L}$, representing, respectively, the energy as measured by an
observer at infinity, the mass of a test particle, the component of the angular momentum along
the axis of symmetry, and the Carter's constant, that can be interpreted as the parallel component of the four momentum
when the particle crosses the equatorial plane of the hole. Sometimes, it will be interesting to use $Q$ instead
$\mathcal{L}$, where $Q=p_\alpha p_\beta K^{\alpha \beta}\geq 0$,
 where ${\bf K}$ is the conformal Killing tensor of the Kerr space-time and
$p^\alpha \equiv dx^\alpha /d\lambda$ is the four momentum of the
test particle (remember that $ Q=\mathcal{L}+ (L-a E)^2$).

An extended summary of geodesics in Kerr space time
can be found in the well known Chandrasekhar book "The
mathematical theory of black holes" (1983) \cite{Chandra}. The
algebraic complexity of the geodesic equation limited the analytical
analysis to particular cases (geodesics in the equatorial
plane, in the extreme Kerr geometry $a=M$, etc..). More recently
E.Teo \cite{Teo} studied spherical light rays out of equatorial
plane finding an analytical expression for the amount of azimuth for
a complete latitudinal oscillation of a spherical orbit (the Wilkins
effect). This work extends an early result of D.C. Wilkins
\cite{Wilkins} devoted to particle spherical orbits in $a=M$ case.

More recently, Drasco and Hughes \cite{DrascoHughes2} have studied
the fundamental orbital frequencies in the "$r$", "$\theta$ " and
"$\phi$" motion of a test particle using Mino time \cite{Mino}. This
work improves an elegant proposal of Schmidt \cite{Schmidt}, based on
the "action angle variables" formalism \cite{G}.

During the last decade, several works have been devoted to the
study of one of the most important problems in gravitation, where
geodesics play an important role, that is, to compute, according to
the General Relativity, the gravitational radiation that one can
expect from different astronomical sources. One kind of these
sources are the so called extreme mass ratio inspiral, EMRI's: These
binary systems are formed as a result of the capture of a compact
stelar remnants by supermassive black holes in the nuclei of
galaxies. The dynamics of such systems are (almost) purely
gravitational, accurately modeled as a point particle of mass $m$,
representing the star, following a trajectory in the background
space-time created by a Kerr Black Hole. If the ratio $m/M$ is
infinitesimal the particle moves along a geodesic. In this system, the particle, having a finite mass, radiates gravitational waves.
In such a case the trajectory deviates from a geodesic, but as the time scale of
the orbital evolution is notably smaller than the typical time scale
of orbits, the particle (approximately) slowly
passes from one geodesic to another, conforming the so called
"adiabatic approximation". Then, radiation reaction effect is
characterized by the time evolution of the parameters $E, L, Q$ of
orbits.

In order to compute this time evolution of parameters different
approaches have been built: using of post Newtonian methods, conservation laws, direct computation of self-force and time-domain
numerical simulations. See for example, among others:
Sago and alt.\cite{Sago} which have obtained an analytic formulae
for the change rates of the energy, angular momentum (using balance
argument) and Carter constant (based partially on a Mino proposal
\cite{Mino} using the radiative field) (see references therein), or
S. Drasco and S.A. Hughes \cite{DrascoHughes1}, that have recently
studied the gravitational wave snapshots of generic extreme mass
ratio inspirals, or the 2007 work "Improved approximated inspirals
of test-bodies into Kerr black holes" of Gair and Glampedakis
\cite{GG}, etc. .

In all these papers adiabatic approximation is implemented, and
thus, an accurately knowledge of orbits in Kerr black hole is
needed. As an interesting example, we focus on the work of different
authors that have contributed to the study of the
gravitational radiation effect on the spherical orbits showing that one
spherical orbit remains spherical after radiation in the adiabatic
approximation (see \cite{Apostolatos}, \cite{Kennefick},\cite{Ryan},
etc..).

In this context, our main goal is to carry out an exhaustive analysis
of orbits of test particle in the outer region of the Kerr
space-time improving our theoretical knowledge, but trying to show
how one orbit is related to each other, in other words, if they can
connect through a continuous change of parameters. We extend the
analytical results on the equatorial plane to the whole outer region
finding the set of values that constants of motion can reach and
showing  the range of coordinates that we expect for an allowed set
of constants of motions of physical null and non-null eccentricity
orbits. This analysis allow us to have a global view of all orbits
that we show in a formal three-dimensional (normalized) parameters
space , ${\hat Q},{\hat L}, $ and ${\hat E}$, where
\begin{equation}
{\hat Q}\equiv \frac{Q}{m^2 M^2}, \ \ \ {\hat L}\equiv
\frac{L}{mM},\ \ \ {\hat E}\equiv \frac{E}{m}. \label{hat}
\end{equation}
In this parameter space each orbit is represented by a point. The
whole set of orbits is confined in a compact volume (see figure
(\ref{estrella}) that partially diverge when ${\hat E}\rightarrow 1$
and vanishes when ${\hat E}<z_a^{-1/2} $ (see equation (\ref{z,x,a})
for $z_{a,b}$ definitions). Spherical orbits are represented by an
upper face (stable) and a lower one (unstable). Slices of this
volume with ${\hat E}=constant$ come first (${\hat E}$ slightly less
than $1$ ) a pseudo-rectangle (upper side to stable, lower side to
unstable, the two bases for orbits on the equatorial plane) , and
after ($z_a^{-1/2} < {\hat E}\leq z_b^{-1/2} $) a pseudo-triangle(
the two sides of spherical orbits joints in one vertex when $z=z_b$
see (\ref{z,x,a})) showing that this triangle vanishes when ${\hat
E}<z_a^{-1/2} $ (see section 4). This representation in the space of
parameters allow us to study the different paths that a particle can
follow going through different geodesics after a continuous change
of constants of motion. As an specific application, we obtain the
necessary and sufficient condition that evolving parameters have to
accomplish to go from one spherical geodesic to another, showing
its limits of application. This result can also be found in
\cite{Apostolatos} and \cite{Kennefick} in different scenarios and
using different techniques.

In section II, III and IV we analyze the geodesics out of the
equatorial plane in Kerr space time using well known results in
literature \cite{Chandra},\cite{Wilkins},\cite{Teo}. In section V
and VI we construct the abstract 3-space of parameters showing the
geometrical locus of spherical and non null eccentricity orbits in that
space, respectively. Some proofs can be find in Appendices I and II.

\vspace{0.5cm}

{\bf First order differential equations for geodesic motion of
massive particles}.

Using the Hamilton-Jacobi method, Carter found the equations of
motion for test particles (see for example \cite{BH}). In the
Boyer-Linquist coordinates these equations are
\begin{eqnarray}
\label{difeq1}\rho^2 \frac{d r}{d \lambda}&=&\pm \{[(r^2 +a^2)E-a
L]^2 -\Delta[m^2 r^2 +(L-a
E)^2+\mathcal{L} \}^{1/2} \label{r}\\
\label{difeq2}\rho^2 \frac{d \theta}{d \lambda}&=&\pm
\{\mathcal{L}-[(m^2-E^2)a^2
+\frac{L^2}{\sin^2\theta}]\cos^2\theta \}^{1/2} \label{theta}\\
\label{difeq3}\rho^2 \frac{d \phi}{d \lambda}&=& -(a E -\frac{L}{\sin^2 \theta})+\frac{a [E(r^2+a^2)-a L]}{\Delta} \label{phi} \\
\label{difeq4}\rho^2 \frac{d t}{d \lambda}&=&-a(a E \sin^2\theta-
L)+\frac{(r^2 + a^2) [E(r^2+a^2)-a L]}{\Delta}. \label{t}
\end{eqnarray}
where
\begin{eqnarray}
\rho^2 &=& r^2+ a^2 cos^2\theta, \label{def rho}\\
\label{def Delta}
\Delta&\equiv&r^2-2 M r +a^2\equiv(r-r_+)(r-r_-),\\
r_{\pm}&=& M{\pm}\sqrt{M^2-a^2},\ \ \ 0\leq r_-\leq M\leq r_+\leq 2
M.
\end{eqnarray}

Without loss of generality, in the intermediate computations we are
going to work with the following dimensionless constants of motion,
often used in the literature (see for example \cite{Chandra}):
\begin{equation}
\eta=\frac{{\hat \mathcal{L}}}{{\hat E}^2},\qquad \xi=\frac{{\hat
L}}{{\hat E}},\qquad z=\frac{1}{{\hat E}^2};
\end{equation}
where ${\hat \mathcal{L}}=\mathcal{L}/m^2M^2$, considering that
$z=0$ for photons, and $z>0$ for massive particles.
In some cases, results exhibit a simpler form in terms of
\begin{equation}\label{mathcalQ}
\mathcal{Q}\equiv {\hat Q}/{\hat E}^2,
\end{equation}
instead of $\eta$. We have to remember that, according with this
definition,
\begin{equation}\label{etamathcalQ}
 \eta=\mathcal{Q}-(\xi-a)^2.
\end{equation}

Replacing these three quantities in the equations of motion and
taking into account that for massive particles $\lambda= \tau/m $,
where $\tau$ is the proper time, the equations become
\begin{eqnarray}
\label{de1p}{\bar \rho}^2 \frac{d x}{d{\bar \tau}}&=&\pm z^{-\frac{1}{2}}{\sqrt R} \\
\label{de2p} {\bar \rho}^2 \frac{d \mu}{d {\bar \tau}}&=&\pm 2 z^{-\frac{1}{2}}{\sqrt \Theta}\\
\label{de3p} {\bar \rho}^2 \frac{d \phi}{d {\bar \tau}}&=& 2
z^{-\frac{1}{2}} [(\frac{1}{1-\mu^2}-\frac{{\bar a}^2}{{\bar
\Delta}})\xi+\frac{4 {\bar a} x}{{\bar \Delta}} ]\\
\label{de4p} {\bar \rho}^2 \frac{d {\bar t}}{d {\bar
\tau}}&=&z^{-\frac{1}{2}}\{{\bar a}[\xi-{\bar a} (1-\mu^2)]+\frac{(4
x^2 + {\bar a}^2)(4 x^2+{\bar a}^2 -{\bar a}\xi)}{{\bar \Delta}}\}.
\end{eqnarray}
where
\begin{eqnarray}
\label{barR} R&=&-x(x-1)(\xi+ \frac{a}{x-1})^2-\frac{\Delta}{4}\{\eta-\frac{4 x^2}{x-1}[z-(z-1)x]\}\\
\label{barTheta} \Theta&=&(1-\mu^2)\eta-\mu^2 \xi^2 -\mu^2 a^2 (z-1)(1-\mu^2),\\
\label{barrho} {\bar \rho}^2&=& 4 x^2 + {\bar a}^2 \mu^2\\
\label{barDelta} {\bar \Delta}&=& 4 x^2 - 4 x + {\bar a}^2\\
\label{x} x&=&\frac{r}{2M},\ \ \  \mu=\cos \theta,\ \ \  {\bar t}=
\frac{t}{2M}, \ \ \ {\bar \tau}= \frac{\tau}{2M}.
\end{eqnarray}

The allowed values of the independent constants of motion $\xi,
\eta$ and $z$ are only limited by conditions $R\geq 0$ and $\Theta
\geq 0$. Equations (\ref{de3p}) and (\ref{de4p}) give us the
variation of $\phi$ and ${\bar t}$ on proper time, once we have
chosen allowed values of these constants.

From now on we will drop the bar symbol from ${\bar a}$, ${\bar
\Delta}$ and ${\bar \rho}$ considering that we will ever work with
non-dimensional quantities.

As it is well known, these equations of motion are valid for all
values of $a$. However, in our work we have differentiated
$a=0$ case (The Schwarzschild space-time) from the $0<a\leq1$ case.

Furthermore, we are interested in the bound geodesics ($z>1$) in the
outer space-time of the hole, that is, $x_+\leq x\leq \infty$.

\section{Theta-motion for bound orbits}

It is well known that, without loss of generality, we can study all
geodesics in Schwarzschild ($a=0$) space-time in the equatorial
plane ($\theta=\pi/2$, $d\theta / d\tau=0$, $\eta=0$).

In order to analyze the whole set of geodesics in Kerr space-time
$0<a\leq0$, the $\theta$-motion equation sets important restrictions on
the $\xi$ and $\eta$ values.

Considering $z\geq 1$ case (we include $z=1$  as a special case)
condition $\Theta\geq 0$ holds if and only if \cite{Chandra}
\begin{eqnarray}\label{etageq0}
\eta&\geq& 0,\\
\label{mumcon}
-\mu_- &\leq& \mu\leq \mu_-,
\end{eqnarray}
in accordance with (\ref{barTheta}), where
\begin{eqnarray}\label{mumdef}
{\mu^2}_- &=& \frac{1}{2\alpha^2} \{{\xi}^2 + \eta + \alpha^2 -
\sqrt{({\xi}^2 + \eta + \alpha^2)^2-4\alpha^2\eta} \}\\
&=&\frac{1}{2\alpha^2} \{{\xi}^2 + \eta + \alpha^2 -
\sqrt{ \xi^4 + (\eta - \alpha^2)^2 +2\xi^2(\eta+ \alpha^2)} \},\nonumber\\
0&\leq& {\mu^2}_- \leq 1 ,\nonumber\\
\alpha^2&=& (z-1)a^2,\nonumber\\
\mu_-&\equiv& +\sqrt{{\mu^2}_-}.
\end{eqnarray}

In the $z=1$ case we find that if $\eta = 0$, then $\mu=0$
($\theta=\pi/2$) or $\xi=0$ and $0\leq\mu^2\leq1$. Moreover, if
$\eta>0$ geodesics may be found.

In the $z>1$ case, $\eta = 0 \Rightarrow
\mu^2(\xi^2+\alpha^2(1-\mu^2)) =0$. It implies: $\mu^2 = 0
\Rightarrow  \theta=\frac{\pi}{2}$, or $\xi^2 +\alpha^2 (1-\mu^2) =
0 \Rightarrow \xi=0$ and $\theta=0,\pi$. Besides, if $\eta> 0$
geodesics exist.

According to (\ref{mumdef}), $|\mu_-|= 1$, if and only if $\xi=0$ and $\eta > \alpha^2$.
It means that only the trajectories with $\xi=0$ can reach
$\theta=0,\pi$, that is, the symmetry axis. This result can be found
not only in particle geodesics case  but also in light rays geodesics (see
\cite{Teo}).

\section{Locally Nonrotating Frame }

There's a condition that arises when we consider that the photon's
or particle's energy measured by  any inertial observer cannot be
negative or zero. We know that if one inertial frame measures
positive energy then all inertial frames must measure positive
energy. We will use an observer who, in some sense, rotates with the
geometry, the so called `locally nonrotating frames LNRF'' (see
\cite{BPT} for definitions). Thus the observer world line is
$r=constant,\theta=constant, \phi=\omega t+constant$ where
$\omega=-g_{\phi t}/g_{\phi phi}$. The orthonormal tetrad carried by
this observer is
\begin{eqnarray}
  {\bf e}_t &=&(\frac{A}{\rho^2 \Delta})^{1/2}\frac{\partial}{\partial t}+\frac{2Mar}{(A \rho^2 \Delta)^{1/2}}\frac{\partial}{\partial \phi},  \
  \ \
  {\bf e}_r = (\frac{\Delta}{\rho^2})^{1/2}\frac{\partial}{\partial
  r},\ \ {\bf e}_\theta =(\frac{1}{\rho^2})^{1/2}\frac{\partial}{\partial
  \theta},\ \ \ {\bf e}_\phi = (\frac{\rho^2}{\Delta\sin^2 \theta })^{1/2}\frac{\partial}{\partial
  \phi}.\nonumber\\
  A&=& (r^2+a^2)^2-a^2 \Delta \sin^2 \theta.
\end{eqnarray}

In this frame the measured energy is
\begin{equation}
E_{LNRF}= - p^\alpha {{\bf e}_t}_\alpha = B^2(E-M w L),
\end{equation}
where $B$ is a non-diverging function and $\omega$ is
\begin{equation}\label{}
w\equiv - \frac{g_{\varphi t}}{g_{\varphi \varphi}}= \frac{4 x a}{M
[(4 x^2+a^2)^2-a^2 \Delta \sin^2 \theta]}.
\end{equation}
It will be interesting to find bounds on $\frac{1}{M \omega}$. These
are
\begin{equation}\label{cotesw}
\frac{1}{M w_m}\geq \frac{1}{M w}\geq\frac{1}{M w_M} >0.
\end{equation}
where
\begin{eqnarray}\label{wMm}
w_M\equiv \frac{4ax}{M (4 x^2+a^2)^2}\nonumber\\
w_m\equiv\frac{a}{M[4x^3 +a^2(x+1)]}.\nonumber
\end{eqnarray}
It is important to remark that
\begin{equation}
w_M(x_+)=w_m(x_+)=\frac{a}{2M(1+\sqrt{1-a^2})}>0\label{wMmxM}.
\end{equation}
The requirement $E_{LNRF}\geq0$ is equivalent to $E- M w L \geq 0$.
Therefore
\begin{eqnarray}
E&>&0\ \ \Rightarrow\ \ \xi\leq \frac{1}{Mw} \leq \frac{1}{M w_m}\ \ \ ,\nonumber\\
E&<&0\ \ \Rightarrow\ \ \xi\geq \frac{1}{Mw}\geq \frac{1}{M w_M}>0
\Rightarrow L<0 \label{lomega}.
\end{eqnarray}

In order to apply this rule out of the equatorial plane, ($\eta \neq
0$), we use the relation (\ref{cotesw}) showing that $\frac{1}{Mw}$ takes values bounded by two
$\theta$-non-depending limits.
\\

In the Schwarzschild case $\omega=0$ and then $E_{LNRF}\geq0
\Leftrightarrow E\geq 0$.

\section{Spherical orbits}

Orbits with constant Boyer-Linquist coordinate  ``$r$''
($x=constant$) require that $R=0$, and $R'=0$, where $R'=d R/d x$.
In the literature one refers to these orbits as 'spherical' or
'circular' orbits indistinctly. The analysis of spherical geodesics
stablishes a basis to classify the whole set of geodesics. We focus our
work on the outer region of Kerr space-time $x_+\leq x$, as we have said
before.

Solving simultaneously $R=0$ and $R'=0$ equations we obtain general
expressions for $\xi_s$ and $\eta_s$, in terms of $x_s$, where $a$
and $z$ are taken as parameters (the subscript $s$ means that these
values correspond to spherical orbits). The solutions for $\xi_s$
are
\begin{equation}
\label{xis12} \xi_{s1,2}=\frac {4{x_s}^2-a^2\pm \Delta_s
\sqrt{2x_s}\sqrt{2(1-z)x_s +z} }{( 2x_s-1) a},
\end{equation}
while the corresponding solutions for $\eta_s$ are
\begin{eqnarray}
\label{etas12} \eta_{s1,2} & = & \frac{4 x_s^2}{ \left( 2\,x_s-1
\right)^{2}{a}^{2}}} \,{  \bigg[ z(2x_s -1)(8 x_s^3-16 x_s^2+8 x_s
-a^2)-\nonumber\\
&-& 4 x_s (4 x_s^3 -8 x_s^2 + 5 x_s - a^2) \mp 2\Delta_s
\sqrt{2x_s}\sqrt{2(1-z)x_s+z}\bigg].
\end{eqnarray}

If $1<z$, the necessary and sufficient condition for these solutions
to exist is:
\begin{equation}
x_s \leq x_{s_{max}}\equiv\frac{z}{2(z-1)}.\label{xsmax}
\end{equation}
Note that $x_{s_{max}}$ is an upper bound to the radius of spherical
orbit. This limit is independent of $a$, as long as
$a\neq 0$.

From now on, we will focus on the following intervals for $x$, $a$
and $z$: $x_+\leq x \leq x_{s_{max}}$, $1\leq z$, $0< a\leq 1$ that
we will generically call $\mathcal{D}$, the domain of variation of these
quantities.

In order to ensure the existence of $\mathcal{D}$ it is necessary that
$x_+\leq x_{s_{max}}$. This implies an upper bound of $z$, that is
\begin{equation}
1\leq z\leq z_{s_{max}},
\end{equation}
where
\begin{equation}
z_{s_{max}}\equiv 1 +\frac{1}{\sqrt{1-a^2}},\label{zmax}
\end{equation}
($a=0, z_{s_{max}}=2$; $a=0.8, z_{s_{max}}=2.66..$; $a\rightarrow1,
z_{s_{max}}\rightarrow \infty$).

If $1\leq z$, the $\theta$-motion analysis has shown that only non
negative values of $\eta$ need to be considered. Then circular
orbits occur when $ \eta_{s1,2}\geq 0 $.

After a tedious calculus we can see that
\begin{equation}
\eta_{s1}(x_s ,a,z)<0 \label{etas1negative}
\end{equation}
in $\mathcal{D}$ (see Appendix I). Therefore spherical orbits in Kerr
space-time ($0 < a \leq 1$) can be found if and only if
$\eta_{s2}\geq0$.

{\bf A}. {\it The  $\{\eta_{s2},\xi_{s2}\}$ solution}.

Firstly, we apply the results of section 3 to this solution. We can
check that
\begin{equation}
\xi_{s2}\leq \frac{ 1}{M\omega_{M}}\leq
\frac{1}{M\omega_{m}},\nonumber
\end{equation}
(if $x=x_+$ we get the equality). Therefore, according to
(\ref{lomega}): No spherical orbits exist if $E<0$ and all spherical
orbits with $E>0$ have $E_{LNRF}>0$.

\vspace{0.5cm}

The analytical complexity of these functions doesn't allow us to have
a complete knowledge of them in a direct way. While $\eta_{s2}$ has
three extrema in $\mathcal{D}$ (every extremum is a maximum, minimum or inflection point depending on the $a$ and $z$ values), $\mathcal{Q}_{s2}$,
defined in (\ref{etamathcalQ}),
\begin{equation}
\mathcal{Q}_{s2}=\eta_{s2}+(a-\xi_{s2})^2
\end{equation}
only has one and shows other interesting properties that we will
discuss below.

The function $R$ can be written as follows
\begin{eqnarray}
  R &=& A( \xi + B)^2 + C(\mathcal{Q}+g),\label{R2}
\end{eqnarray}
where $A,B,C,g$  are
\begin{equation}
  A =\frac{a^2}{4}, \ \ \ B(x,a) = -\frac{4x^2+a^2}{a}, \ \ \ C(x,a) = -\frac{\Delta}{4},\ \ \
  g(x,z)=4zx^2.
\end{equation}

Thus $R$ must be considered as a function of $x,a, C^i$, i.e. $R(x,a, C^i)$, where $C^i
\equiv\{\xi,\mathcal{Q},z\}$. According to this we define
\begin{eqnarray}
  R' &\equiv&\frac{\partial R}{\partial x},  \ \ \ R''\equiv\frac{\partial R'}{\partial x}\label{Rp,Rpp}\\
  B' &\equiv&\frac{d B}{d x},\  \ \ C' \equiv\frac{d C}{d x}, \ \ \ g' \equiv\frac{d g}{d x} \label{dBC}.
\end{eqnarray}
For each set of allowed values of $C^i$ we can find a solution
$x=x(\tau,C^i,a)$ solving equation (\ref{de1p}). Thus
$R=R(x(\tau,C^i,a),C^j,a)$.

We are now interested in a family of solutions such that the
parameters vary according to a specific law, defined by the functions
$C^i(s)$, assuming that for each value of $s$ we obtain a geodesic
\cite{Kennefick}. The solution of (\ref{de1p}) now takes the form
$x=x(\tau,C^i(s),a)$, and thus
\begin{eqnarray}
R=R(x(\tau,C^i(s),a),C^j(s),a).\label{R2b}
\end{eqnarray}
If we produce a small variation $\delta s$ then the quantities $R$
and $R'$ vary according to
\begin{eqnarray}
 \frac{\delta R}{\delta s} &=& R' {\dot x} + \stackrel{i=1,3}{\Sigma} \frac{\partial R}{\partial C^i}{\dot C^i} \label{Rc}\\
 \frac{\delta R'}{\delta s}&=&R'' {\dot x} + \stackrel{i=1,3}{\Sigma} \frac{\partial R'}{\partial C^i}{\dot
 C^i} \label{Rd}.
\end{eqnarray}
where ${\dot x}\equiv {\partial x}/{\partial s}$ and ${\dot
C^i}\equiv {\partial C^i}/{\partial s}$.

We apply this results in order to investigate the  properties of the
first derivatives of $\xi_{s2} (x_s,a,z)$ and $\mathcal{Q}_{s2}
(x_s,a,z)$.

Spherical orbits imply $x=s\equiv x_s$. This means ${\dot x}=1$,
thus, $\xi(s\equiv x_s)=\xi_{s2}(x_s,z,a)$, $\mathcal{Q}(s\equiv
x_s)=\mathcal{Q}_{s2}(x_s,z,a)=\eta_{s2}(x_s,z,a)+(a-\xi_{s2}(x_s,z,a))^2$
and ${\dot z}=0$. For every allowed value of $s\equiv x_s$ we have
spherical orbits, therefore $R(x_s, \xi_{s2},\mathcal{Q}_{s2}, z,
a)=R'(x_s, \xi_{s2},\mathcal{Q}_{s2}, z, a)=0$. As we go from one
spherical orbit to another varying $s$, hence ${\dot R}(x_s,
\xi_{s2},\eta_{s2}, z, a)={\dot R'}(x_s, \xi_{s2},\eta_{s2}, z,
a)=0$. Using these results on (\ref{Rp,Rpp}), (\ref{Rc}) and
(\ref{Rd}) we obtain
\begin{eqnarray}
0 &=& 2A(\xi_{s2}+B_s)B'_s+ C'_s(\mathcal{Q}_{s2}+g_s)+C_s g'_s ,\label{cero}\\
0 &=& 2 A (\xi_{s2}+B_s)\xi'_{s2} + C_s\mathcal{Q}'_{s2} \label{una}\\
0 &=& R''_s+ 2 A B'_s \xi'_{s2}+\nonumber\\
  &+& C'_s\mathcal{Q}'_{s2} \label{dos},
\end{eqnarray}
where for example $B_s\equiv B(x=x_s)$, and ${\dot \xi_{s2}} \equiv
\partial \xi_{s2}/\partial x_s \equiv \xi'_{s2}$, in accordance with our previous notation.
\vspace{0.5cm}

{\bf Theorem I}:  $\xi'_s(x_e(a,z),a,z)=0$ if and only if
$\mathcal{Q}'_s (x_e(a,z),a,z)=0$, where $x_s= x_e(z,a)$ is a
solution of $\xi'_s(x_s,a,z)=\mathcal{Q}'_s (x_e(a,z),a,z)=0$. In
other words, there where $\xi_s (x_s,a,z)$ has an extremum, the
function $\mathcal{Q}_s (x_s,a,z)$ must have an extremum, and
vice versa.

Proof: Considering that $\xi_{s2}$ and $\mathcal{Q}_{s2}$ satisfy
$R'(\xi_{s2},\eta_{s2}, x_s,a,z)=0$, then, according to (\ref{una}),
if $\xi'_{s2} =0$ then $\mathcal{Q}'_{s2}=0$ since $C_s=-\Delta_s/4
\leq 0$ ($x_+\leq x_s \leq x_{s_{max}}$). Conversely, if
$\mathcal{Q}'_{s2}=0$ it implies that $\xi'_{s2}=0$ since the other
possibility $\xi_{s2}=-B_s$, $g'_s=0$ and $\mathcal{Q}_{s2}=-g_s$
(to ensure $R=R'=0$)is impossible due to $g_s>0$ and
$\mathcal{Q}_{s2}$ must be greater than zero.

In Appendix III we have proved that this extremum exists and is
unique. We call it $x_{e}(a,z)$ but we are not able to found his analytical
expression.

From equation $\xi'_{s2}=0$ we can obtain
\begin{eqnarray}
z_e &=&\frac{8 x_e}{(32 x_e^3 -36 x_e^2+ 12 x_e -a^2)^2}[2 x_e(2 x_e -1)(32 x_e^3 - 44 x_e^2 + 20 x_e - 3 a^2)-\nonumber\\
&-&\Delta_e^\frac{3}{2} \sqrt{2x_e}].\label{ze}
\end{eqnarray}
Using this expression we find
\begin{eqnarray}
  \xi_{e2}(x_e,a) &=&  \xi_{s2}(x_s=x_e,a,z=z_e)\label{xize}\\
  \mathcal{Q}_{e2}(x_e,a)&=&\mathcal{Q}_{s2}(x_s=x_e,a,z=z_e)\label{Qze}.
\end{eqnarray}
where the functions $\xi_{e2}(x_e,a)$, $z_e(x_e,a)$ and
$\mathcal{Q}_{e2}(x_e,a)$ represent the values of minimum of
$\xi_{s2}$, maximum of $\mathcal{Q}_{s2}$ and $z_e$ when the extremum
is in $x_e$.

{\bf Theorem II}: The function
$R''(\xi_{s2},\mathcal{Q}_{s2},x_s,a,z)$ vanishes if and only if t
$\xi'_{s2}(x_s,a,z)=\mathcal{Q}_{s2} (x_s,a,z)=0$.

Proof: The right to left implication is obvious using (\ref{dos}). The left to right implication
 can be proved considering that the determinant of unknowns
$\xi'_{s2}$ and $\mathcal{Q}'_{s2}$ in the homogeneous system of two
equations (\ref{una}) and (\ref{dos}), considering that now $R''=0$,
is
\begin{equation}
Det =2 A_s (\xi_{s2} +B_s) C'_s - 2A_s B'_s
C_s=\frac{\Delta_s}{2}\sqrt{-2x_s(2x_s z-2x_s-z)}.
\end{equation}
It is clear that $Det$ only vanishes in $x=x_+$ and $x=x_{s_{max}}$.
Then the only solution is
$\xi'_{s2}(x_s,a,z)=\mathcal{Q}'_{s2}(x_s,a,z)=0$ in $x_+ < x_s <
x_{s_{max}}$.

As it is well known stable (unstable) spherical orbits occur when
$R''(x,a,\mathcal{Q}_{s2}, \xi_{s2},z)<0 (>0)$.

Theorem II implies that $R''=0$ only where
$\mathcal{Q}'_{s2}(x_s,a,z)=\xi'_{s2}(x_s,a,z)=0$, that is, where
$x=x_{e}(a,z)$. It is not difficult to prove that spherical orbits
are stable when $x_s>x_{e}(a,z)$. Conversely, if $x_s<x_{e}$ they
are unstable.

{\bf Theorem III}:The extremum of $\xi_{s2}$ and $\mathcal{Q}_{s2}$
are maximum (minimum) and minimum (maximum) respectively in $\mathcal{D}$.

From (\ref{una}), and considering the point where
$\xi'_{s2}=\mathcal{Q}'_{s2}=0$ we obtain
\begin{equation}\label{xiQpp}
\frac{\xi''_{s2}}{\mathcal{Q}_{s2}''}=-\frac{C_s}{2A_s(\xi_{s2}+B_s)}<0
.
\end{equation}
Numerical calculations show that this point is always a maximum of
$\mathcal{Q}_{s2}$ and a minimum of $\xi_{s2}$.

The values of these two functions in $x_+$ and $x_{s_{max}}$ are
\begin{eqnarray}\label{}
\mathcal{Q}_{s2}(x_+,a,z)&=&-(1+\sqrt{1-a^2})z< 0,\nonumber \\
\mathcal{Q}_{s2}(x_{s_{max}},a,z)&=&=- \frac{z^2}{z-1} [z(1-a^2)+a^2] < 0,\nonumber\\
\xi_{s2}(x_+,a,z)&=& \frac{2(1+\sqrt{1-a^2})}{a}>0,\label{xis2x+} \\
\xi_{s2}(x_{s_{max}},a,z)&=& \frac{z}{a(z-1)}[z(1-a^2)+a^2]+a
>0.\label{xis2xmax}
\end{eqnarray}

{\bf B} $\xi_{s2}=\xi_{s2}'=0$.

It will be important to find the functions $z_c(a)$ and $x_{ec}(a)$
such that if $z>z_c$ then $\xi_{s2}(x_s,a,z)>0$, where $x_{ec}$ is
the point where $\xi_{s2}=\xi_{s2}'=0$, for any value of $a$. To
do this we proceed as follows: First, the system of equations
$\xi_{s2}=\xi_{s2}'=0$ is equivalent to
\begin{eqnarray}\label{axi}
a&=&2x_{ec}^{1/2} \left[-
x_{ec}+1+\frac{1}{9}\left(A(x_{ec})^{1/3}-A(x_{ec})^{-1/3}+\sqrt{2}\right)^2\right]^{1/2},\\
\label{zc} z_c &=& \frac{1}{2}\,{\frac {\left( 4\,{x_{ec}}^{2}+a^2
\right) \left( 8\,{x_{ec}}^{3}- 12\,{x_{ec}}^{2} +2\,a^2 x_{e3c}+a^2
\right) }{ \left( 4\,{x_{ec}}^{2}-4\,x_{e3c}+a^2 \right)
^{2}x_{ec}}},
\end{eqnarray}
where
\begin{equation}\label{zxi}
A(x_{ec})=11\sqrt{2}-27\sqrt{2} x_{ec}+
3\sqrt{27-132x_{ec}+162x_{ec}^2}.
\end{equation}

\begin{figure}[!htb]
\includegraphics[width=0.5\textwidth]{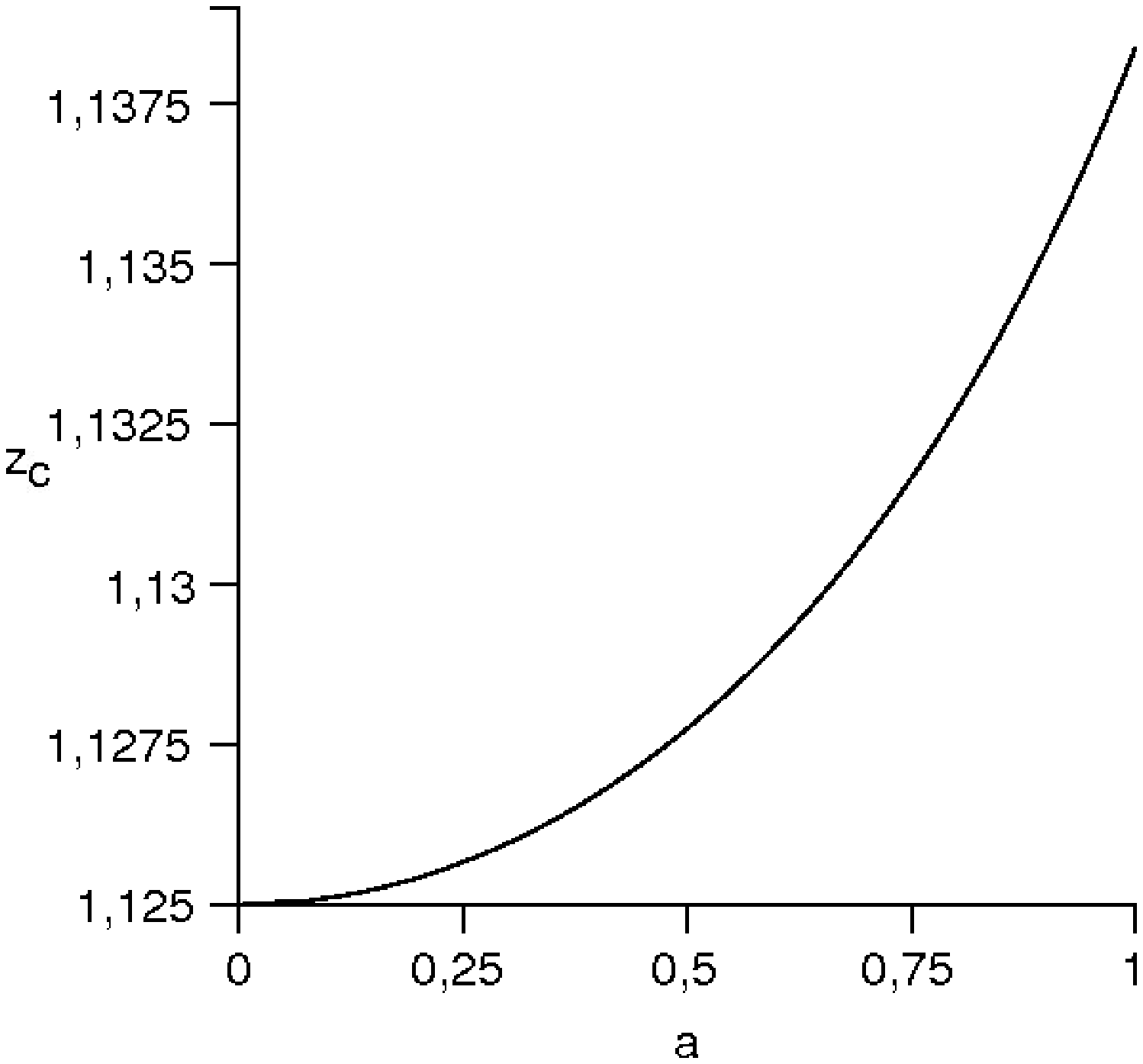}
\includegraphics[width=0.5\textwidth]{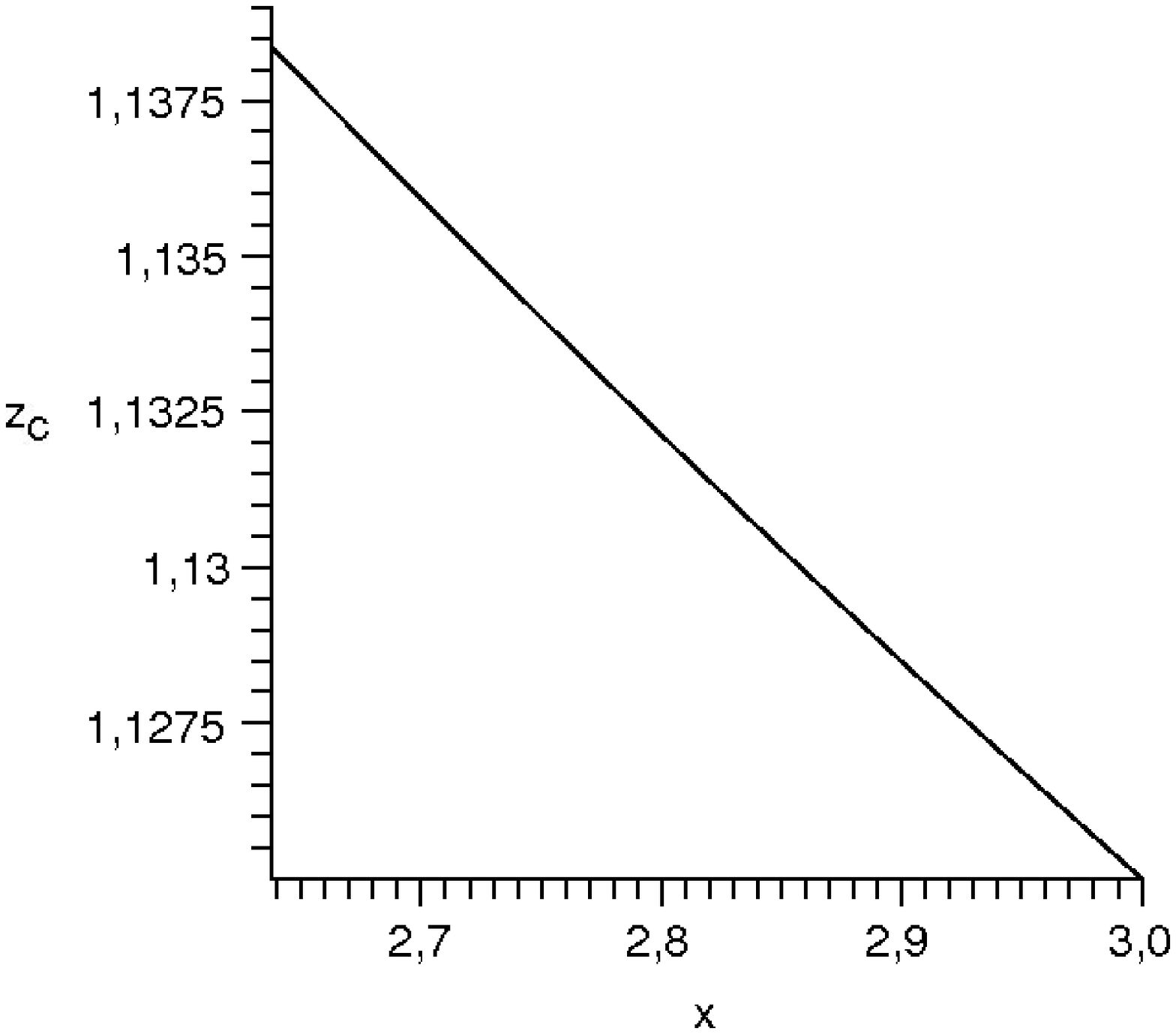}
\caption{\label{z-a--xi=xip=0}Plot of $z_c(a)$ and $z_c(x_{ec})$}
\end{figure}

If we use (\ref{axi}) in (\ref{zc}) we obtain a function
$z_c(x_{ec})$ that combined with (\ref{axi}) is a parametric
expression of $z_c(a)$ (see figure (\ref{z-a--xi=xip=0}),a), that
is, for every $a$ we find $z_c$. Then, using (\ref{zxi}) (see
figure (\ref{z-a--xi=xip=0}), b) we can find $x_{ec}$ value.

Numerical examples can be done: $a=0, x_{ec}=3,z_c=1.125$;
 $a=0.8,x_{ec}=2.784,z_c=1.133$ and
$a=1,x_{ec}=2.637255281,z_c=1.138373645$.

As an interesting application of this result, we can prove that $d\phi/d\tau \geq 0$ for all spherical orbits provided  $z\geq z_c$. According to (\ref{de3p})
\begin{equation}
\frac{z^{\frac{1}{2}} {\bar \rho}^2}{2} \frac{d \phi}{d {\bar \tau}}=
(\frac{1}{1-\mu^2}-\frac{a^2}{\Delta})\xi_{s2} +\frac{4 a x}{\Delta} \geq  (1-\frac{a^2}{\Delta})\xi_{s2} +\frac{4 a x}{\Delta}.\nonumber
\end{equation}
But
\begin{equation}
(1-\frac{a^2}{\Delta})\xi_{s2} +\frac{4 a x}{\Delta}-\xi_{s2} = {\frac {a \left( 1+\sqrt {2}\sqrt {- \left( -2\,x+2\,zx-z \right) x}\right) }{2\,x-1}}> 0. \nonumber
\end{equation}
Therefore
\begin{equation}
\frac{z^{\frac{1}{2}} {\bar \rho}^2}{2} \frac{d \phi}{d {\bar \tau}}=(\frac{1}{1-\mu^2}-\frac{a^2}{\Delta})\xi_{s2} +\frac{4 a x}{\Delta} > \xi_{s2}.
\end{equation}

Thus, we can conclude that:

{\bf Theorem IV}. For every value of $a$,
there exists $z_c$, defined by (\ref{zc}) and (\ref{axi}), such that
if $z\geq z_c$ then $\xi_{s2}\geq 0$ in $\mathcal{D}$ and thus $d\phi/d\tau > 0$, i.e. the spherical orbits co rotate with the hole.

{\bf C}
$\mathcal{Q}_{s2}-(a-\xi_{s2})^2=\mathcal{Q}'_{s2}-[(a-\xi_{s2})^2]'=0$.

In order to improve our knowledge of $x_{e}$, we can find where both
$\mathcal{Q}_{s2}=(a-\xi_{s2})^2$ and
$\mathcal{Q}'_{s2}=[(a-\xi_{s2})^2]'=0$, or alternatively
$\eta_{s2}={\eta_{s2}}'=0$. These conditions are equivalent to
\begin{eqnarray}
z_{a,b} &=& \frac{3x_{ea,b}}{3 x_{ea,b}-1}, \label{z,x,a}\\
x_{ea,b} &=&\frac{1}{2} \{ 3+Z_2 \mp
[(3-Z_1)(3+Z_1+2Z_2)]^\frac{1}{2} \}.\label{x3ab}
\end{eqnarray}
where
\begin{eqnarray}
Z_1 &=& 1+(1-a^2)^\frac{1}{3}[(1+a)^\frac{1}{3}+(1-a)^\frac{1}{3}],\nonumber \\
Z_2 &=& (3a^2+{Z_1}^2)^\frac{1}{2}.\nonumber
\end{eqnarray}
Functions $x_{ea,b}$ where first presented in \cite{BPT} (see
equation (2.21)in that paper). We now show some values of these
variables:
\begin{itemize}
\item When $a=1$ then $z_a=3$ and $z_b=1.080$ while $x_{ea}=1/2$ and
$x_{eb}=9/2$,
\item When $a=0.8$ then $z_a=1.298$ and $z_b=1.086$
while $x_{ea}=1.453$ and $x_{eb}=4.216$,
\item When $a\rightarrow0$ then $z_a, z_b\rightarrow 9/8$ while $x_{ea}, x_{eb}\rightarrow 3$.
\end{itemize}
Moreover, $x_{ea,b}$ give us the range of variations of $x_{e}$,
that is $x_{eb}(a)\leq x_{e}(z,a)\leq x_{ea}(a)$ obtaining
non-negative values of $\eta_{s2}$.

\begin{figure}[!htb]
\centering
\includegraphics[width=0.5\textwidth]{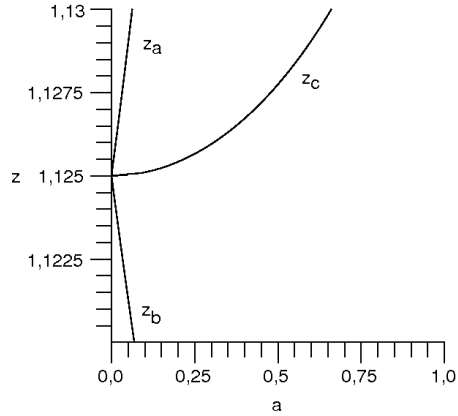}
\caption{\label{z1z2zazbzc} Dependence on $a$
of $z_b, z_c$ and $z_a$.}
\end{figure}

Finally, we can verify that $1<z_b <z_a$, see fig (\ref{z1z2zazbzc}).

The behavior of these functions near $a=0_+$  is
\begin{eqnarray}
z_b &=& \frac{9}{8}-\frac{\sqrt{6}}{32} a + O( {a}^{2}) \nonumber\\
z_c &=& \frac{9}{8}+ \frac{1}{48} a^2+ O({a}^{3})\nonumber\\
z_a &=&\frac{9}{8}+\frac{\sqrt{6}}{32} a + O ({a}^{2})\nonumber.
\end{eqnarray}
this implies that $z_b \leq z_c \leq z_a$ (the equality holds in the
limit $a\rightarrow0$).

{\bf D} {\it Classification}.

\begin{figure}[!htb]
\includegraphics[width=0.24\textwidth]{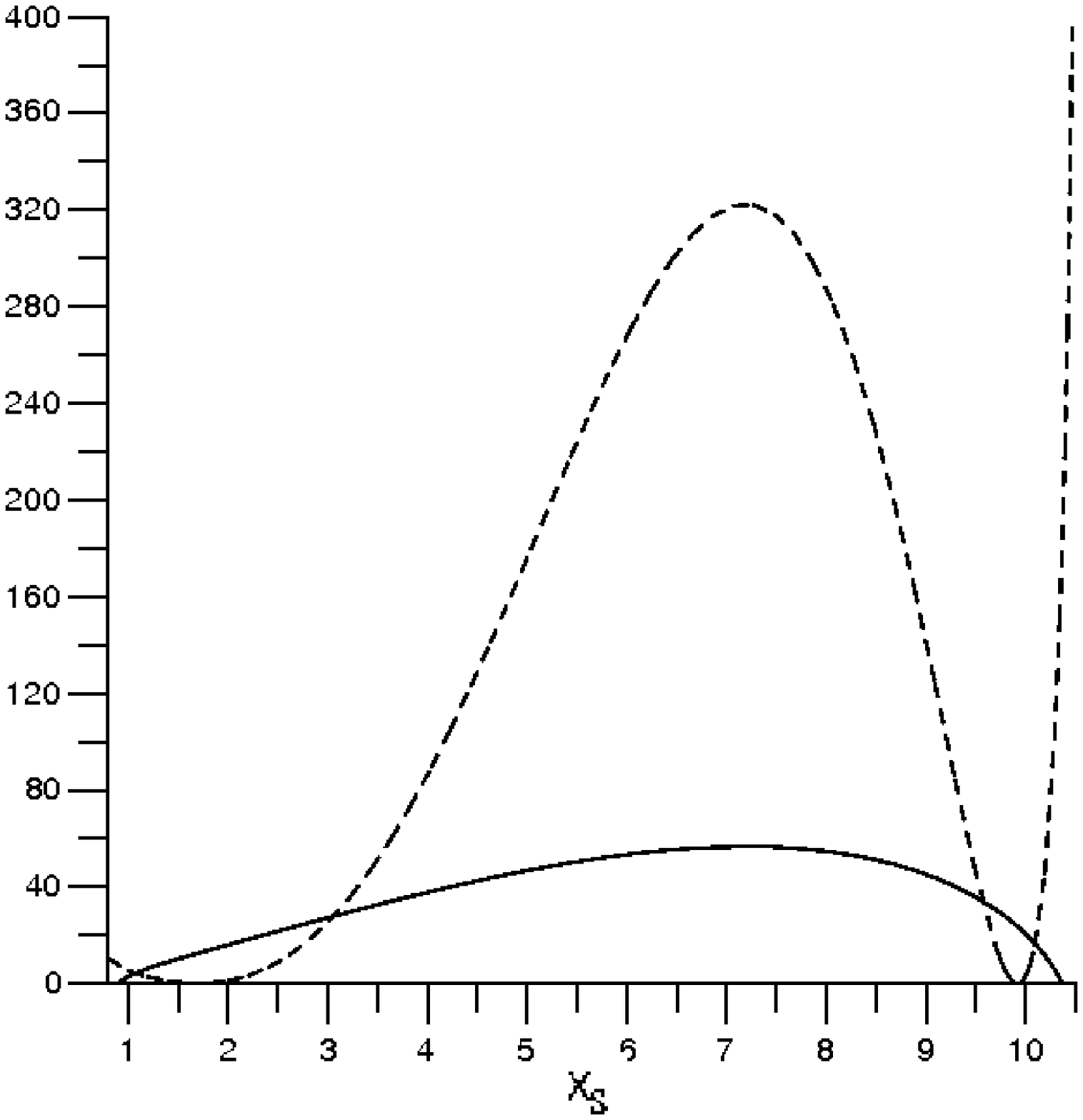}
\includegraphics[width=0.24\textwidth]{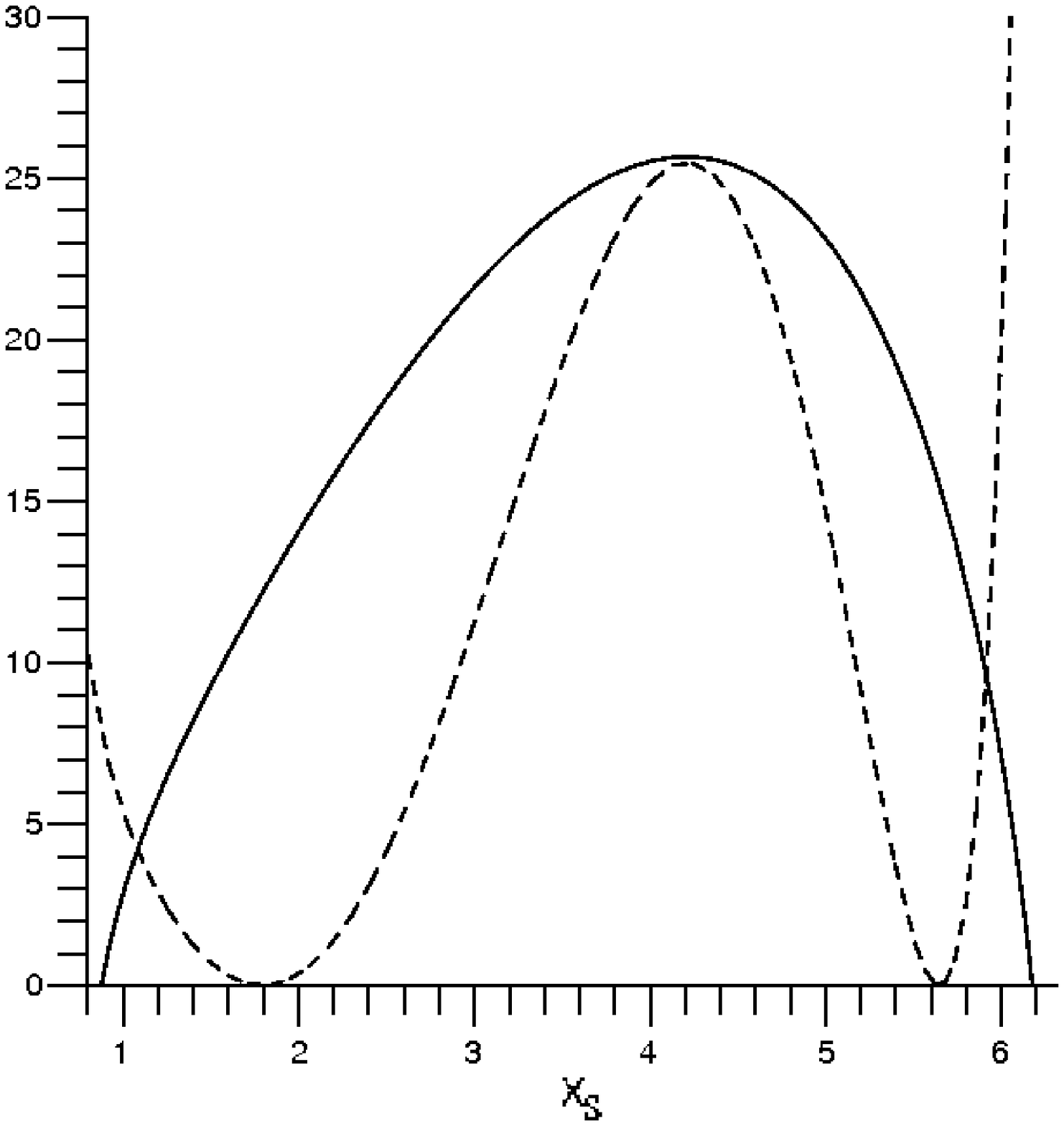}
\includegraphics[width=0.24\textwidth]{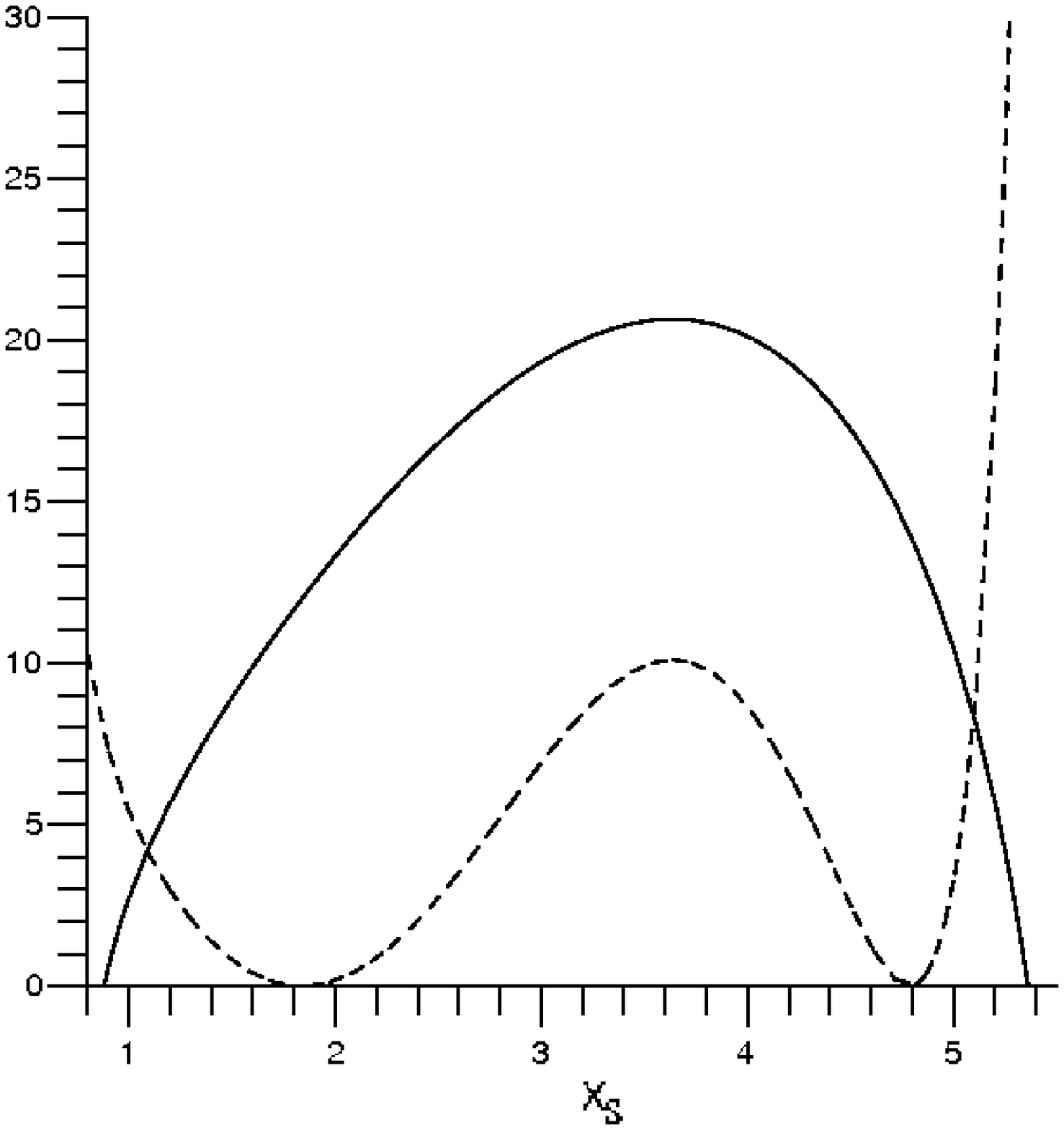}
\includegraphics[width=0.24\textwidth]{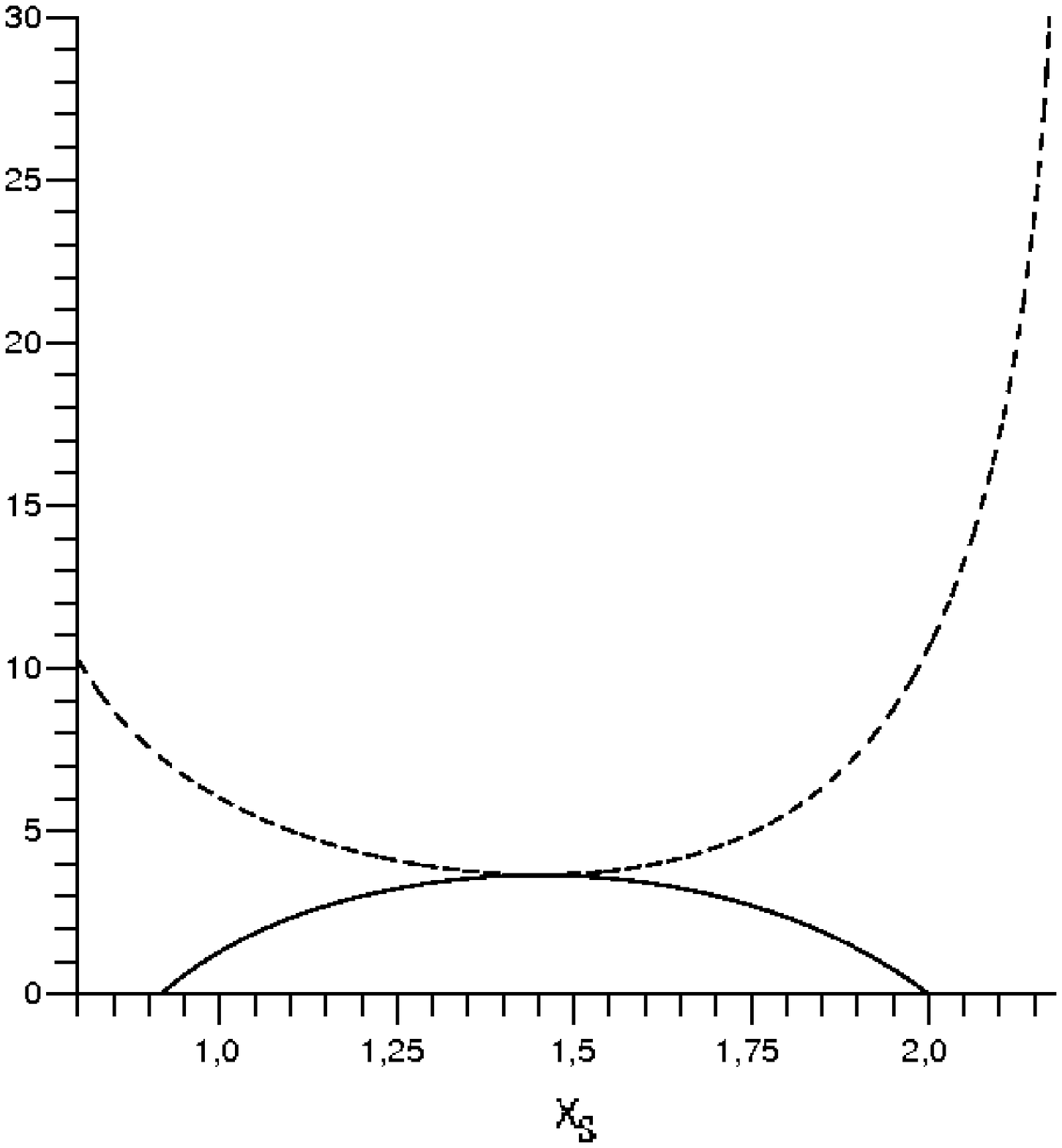}
\caption{\label{scenarios}Plot of $Q_{s2}$ (continuous line) and $(a-\xi_{s2})^2$ (dashed line) corresponding to $I;\
II;\ III;\ IV$ cases}
\end{figure}

Until now we have studied the functions $\mathcal{Q}_{s2}$ and
$\xi_{s2}$. We conclude this subsection analyzing under which
conditions  $\mathcal{Q}_{s2}\geq (a-\xi_{s2})^2$ ($\Leftrightarrow\eta_{s2}\geq 0$) holds.
We call $x_i$, $i,j,..\equiv\{1,2,...\}$, $x_i\leq x_{i+1}$, the
points where $\mathcal{Q}_{s2}=(a-\xi_{s2})^2$. We will depict these two functions $\mathcal{Q}_{s2}$ and $(a-\xi_{s2})^2$
in four possible scenarios (see figure \ref{scenarios}):

\begin{itemize}
\item I)$1<z<z_b$. Condition only holds in
two disconnected intervals, $x_+<x_1\leq x\leq x_2<x_e $ for the unstable spherical orbits (in which $Q_{s2}$ is an increasing function) and $<x_e<x_3\leq x\leq
x_4<x_{s_{max}}$ that corresponds to the stable ones (in which $Q_{s2}$ is a decreasing function),
\item II)$z=z_b$. The two parts are just connected in $x_{e}$,
that is $x_2=x_e=x_3$ of the previous case, and we can continuously
go from the stable to the non-stable orbits,
\item III) $z_b<z<z_a$. The condition holds in the whole interval $x_+<x_1\leq x \leq x_4 <
x_{s_{max}}$ containing stable and unstable orbits (now $x_2$ and
$x_3$ doesn't exist).
\item IV) $z=z_a$ We only have one spherical geodesic in $x_e$.
\end{itemize}

\begin{figure}[!htb]
\centering
\includegraphics[width=0.5\textwidth]{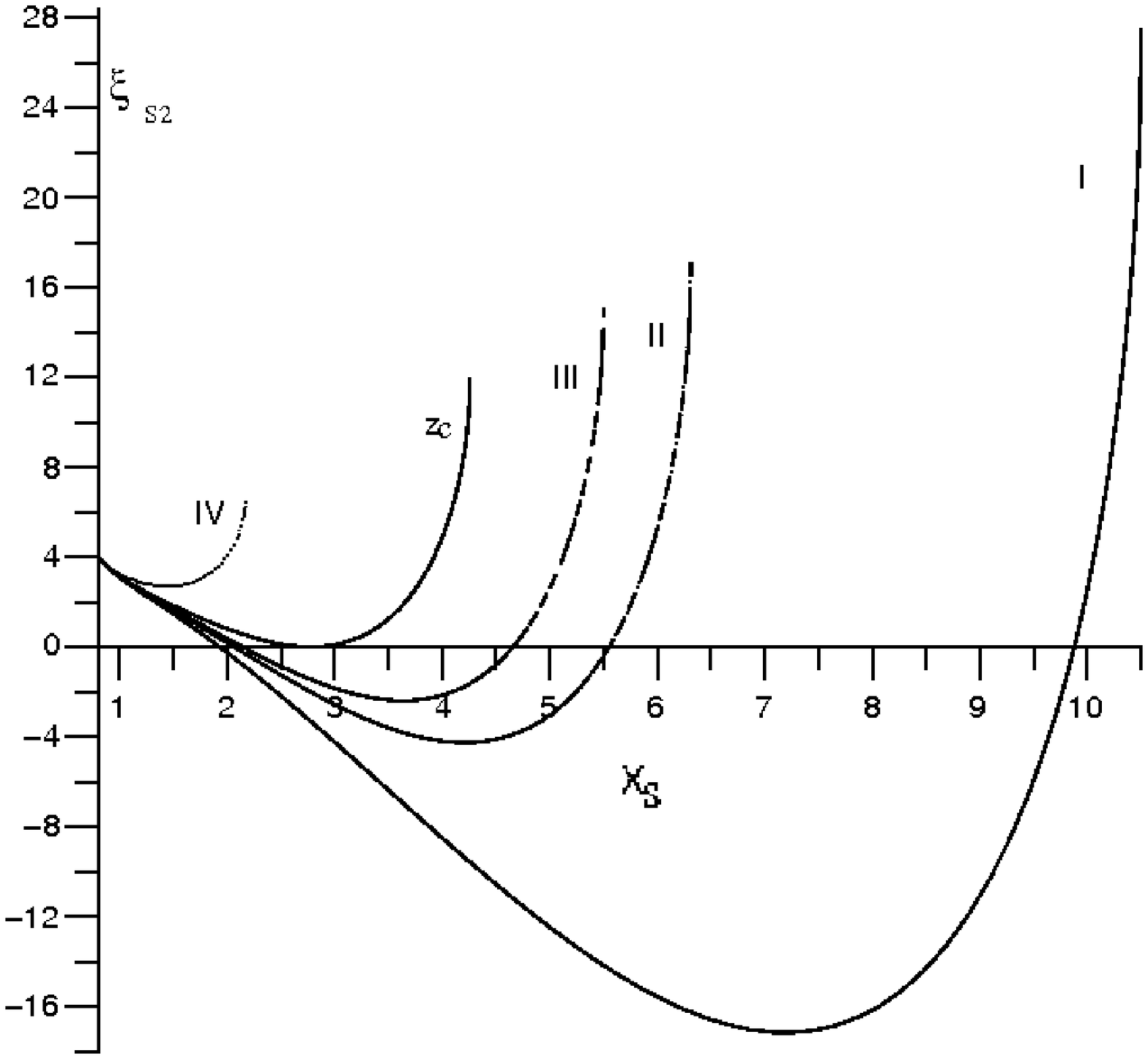}
\caption{\label{scenarios_xis2}}
\end{figure}

In figure \ref{scenarios_xis2},
we can see $\xi_{s2}$ in these cases, including $z=z_c$.

{\bf E} {\it Equatorial Plane}.

As it is well known, a necessary condition to have orbits lying
entirely in the equatorial plane is $\eta=\mathcal{Q}-(\xi-a)^2=0$.
For spherical orbits the condition becomes $\eta_{s2}=0$.

Conditions $R=R'=0$ are now ($\eta=0$) equivalent to
\begin{eqnarray}
z_{s\pm} &=& 4\,{\frac { \left[
16\,{x_s}^{5}-56\,{x_s}^{4}+64\,{x_s}^{3}-6 \left({a}^{2} +4
\right) {x_s}^{2}+ 5\,{a}^{2}x_s \pm a \Delta_s \sqrt
{2x_s}\right] x_s}{ \left(8\,{x_s}^{3} -16\,{x_s}^{2}
+8\,x_s-{a}^{2}\right) ^{2}}}\nonumber
 \\
\xi_{s\pm} &=&\frac {-a(12\,{x_s}^{2}-8\,x_s +{a}^{2})\mp 2\, \Delta_s x_s \sqrt {2x_s}}{8\,{x_s}^{3}
-16\,{x_s}^{2} +8\,x_s-{a}^{2}}\label{xieq}\\
\mathcal{Q}_{s\pm}&=&(\xi_{s\pm}-a)^2.
\end{eqnarray}
These equations are equivalents to (2.12) and (2.13) of \cite{BPT}.
\begin{figure}[!htb]
\centering
\includegraphics[width=0.5\textwidth]{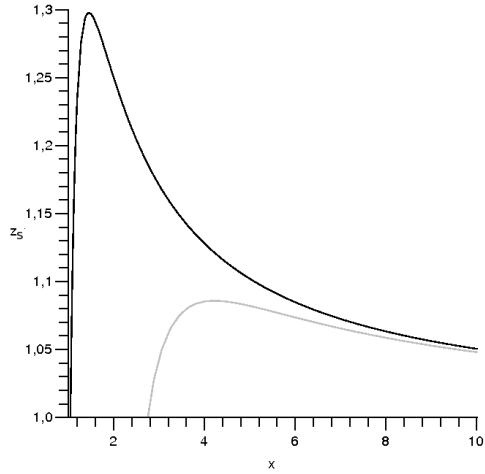}
\caption{\label{zfxa}Graphic $z(x_s)$, $\eta=0$, $a=0.8$}
\end{figure}

Each of $z_{s\pm}$ solutions has a maximum in the interval $1\leq z
\leq \infty$, that corresponds with $z_{_\pm M}=z_{a,b}$, $x_{\pm
M}=x_{ea,b}$ defined in (\ref{x3ab}). Beyond these points they
decay asymptotically to one, as we can see in figure (\ref{zfxa}).

As you can see in the last figure, if
\begin{itemize}
\item $1<z<z_b$ we have four spherical orbits.
\item $z=z_b$ we have three, one of them in $x_{eb}$ such that $\{z_b,x_{eb}, a\}$ are solutions of (\ref{z,x,a}),(\ref{x3ab}) equations ($z_b<z_a$),
\item $z_b<z<z_a$ we have two,
\item $z=z_a$ we have the last one, such that $\{z_a,x_{ea}, a\}$ are
its solutions.
\end{itemize}

A specially simple expression of the function $z_{s\pm}(x_s,a)$ can
be obtained for $z=1$, and $z=9/8$ values, that are well known in
the literature.

\begin{eqnarray}
a(z=1)=\sqrt{4(x_s-2 \sqrt{2x_s}+2)x_s}=2
\sqrt{x_s(\sqrt{x_s}-\sqrt{2})^2}.\label{a2xz1}
\end{eqnarray}

\begin{equation}\label{}
a(z=9/8)=\frac{2\sqrt{2x_s}}{9} \sqrt{-7{x_s}^2+ 18x_s + 81 \pm
4(9-x_s)\sqrt{x(9-2x_s)}}.\label{a12z98}
\end{equation}
(See \cite{BPT} and \cite{Chandra} for details).

{\bf F} {\it Schwarzschild circular orbits}.

Conditions $R=R'=0$ imply
\begin{eqnarray}
  \xi_{s\pm} &=&\pm \frac{\sqrt{2} x_s^\frac{3}{2}}{x_s-1}\label{schxi} \\
  z_s &=& \frac{(2x_s-3)x_s}{2(x_s-1)^2}.\label{schz}\\
  \mathcal{Q}_{s\pm}&=&\xi^2_{s\pm}.
\end{eqnarray}

Functions $\xi_{s-}$ and $z_s$ have a maximum,
${\xi_{s-}}_M=-\frac{3\sqrt{6}}{2}$, ${z_{s}}_M=9/8$ respectively,
in $x_s=3$, and $\xi_{s+}$ a minimum,
${\xi_{s+}}m=\frac{3\sqrt{6}}{2}$.

Stable circular orbits can be found when $3< x_s <\infty$, and
unstable $2\leq x<3$, where  $x_s=2, z_s=1, \xi_{s\pm}=\pm 4$
corresponds to the inner unstable circular orbit. (See \cite{BPT}
and \cite{Chandra} for details).

\section{Spherical orbits in the ${\hat Q}-{\hat L}-{\hat E}$ space}

In this section we are interested in representing ``outer''
spherical orbits in a  ${\hat Q}$, ${\hat L}$ and ${\hat E}$
(\ref{hat}) formal three-space,($\ \ {\hat Q}=z^{-1} \mathcal{Q}$,
${\hat L}\equiv z^{-1/2}\xi$ and  ${\hat E}\equiv z^{-1/2}$),
analyzing their properties, as we have mentioned before. This task can be achieved due to the
special shape of the geodesic equation that, at the same time,
derives from the existence of two killing vector fields and one
conformal killing 2-tensor. As we have seen $\xi_{s2}$ and
$\mathcal{Q}_{s2}$ depend exclusively on $x=r/2m$ coordinate. This
fact allow us to construct a formal three space in such a way that
the whole set of spherical orbits will be represented by a
2-surface, say $\Sigma$, in this space. In the next chapter we will
see that orbits with non null eccentricity can be represented in the
same formal space. Therefore, we will achieve a global and careful
(with accurate analytical control) representation of the whole set
of orbits in Kerr space-time.

{\bf A}. Each point of this space is characterized by their
coordinates $\mathcal{X}^i \equiv \{{\hat Q},{\hat L}, {\hat E}\}$.

The geometrical locus of the spherical orbits in this space is a
2-surface $\Sigma$ defined by:
\begin{eqnarray}
{\hat Q} &=& z^{-1} \mathcal{Q}_{s2}(x_s,z,a),\nonumber\\
{\hat L} &=& z^{-\frac{1}{2}}\xi_{s2}(x_s,z,a),\nonumber\\
{\hat E} &=& z^{-\frac{1}{2}}\label{QLE},
\end{eqnarray}
limited by
\begin{equation}\label{Qequ}
 {\hat Q}= ({\hat L}-a {\hat E})^2,
\end{equation}
that is, the surface representing the equatorial plane in this
space ($\eta=0$), and $0<{\hat E}\leq 1$. Finally $x_s$ and $z$
($z={\hat E}^{-2}$) have to be considered the two parameters
constrained by
\begin{eqnarray}
 x_+(a) &\leq & x_s\leq x_{s_{max}}(z),\nonumber\\
 1&<& z \leq z_a(a)\nonumber.
\end{eqnarray}
where $z_a$ is defined in (\ref{z,x,a}). We will look at this representation $a$ ($0<a\leq 1$) as a constant
(for example $a=0.8$), discussing its variation in the results and
including $z=1$ as a limiting case.

Moreover, this representation is not sensitive to initial
conditions. Then, one point represents spherical orbits but, in some
circumstances, the same point represents the associated geodesics
(see next section for definition).

It is well known (see \cite{Struik} for example) that singular
points occur when two minors of the matrix
\begin{equation}
\left(
  \begin{array}{ccc}
    \frac{\partial {\hat Q}}{\partial x_s} &\frac{\partial {\hat L}}{\partial x_s} & \frac{\partial {\hat E}}{\partial x_s}=0 \\
    \frac{\partial {\hat Q}}{\partial z}&\frac{\partial {\hat L}}{\partial z} & \frac{\partial {\hat E}}{\partial z} \\
  \end{array}
\right)
\end{equation}
 vanishes. This condition holds if and only if $\xi'_{s2}=\mathcal{Q}'_{s2}=0$. The solution of these equations is $x_s =x_{e}(z,a)$. Replacing this solution in (\ref{QLE}) we obtain the set of singular points of $\Sigma$.

{\bf B}.

\begin{figure}[!htb]
\includegraphics[width=1\textwidth]{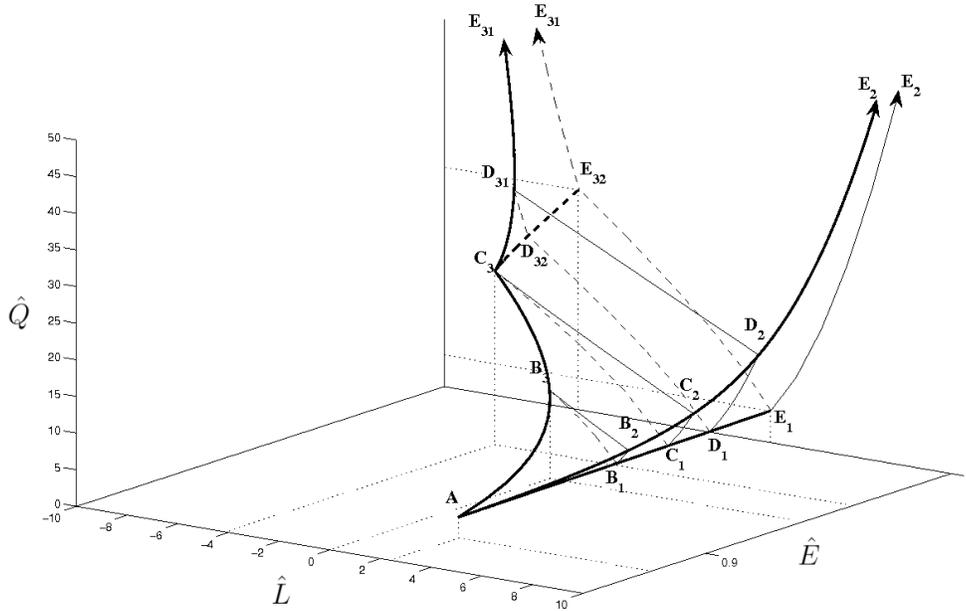}
\caption{\label{estrella} The geometrical locus of spherical orbits
in Kerr space time. Slices are made for $I,z=1.05; II, z=z_b=1.086; III, z=z_c=1.133; IV, z=z_a=1.298$}
\end{figure}

Spherical orbits can be represented using a parametric plot of
equations (\ref{QLE}). The range of variation of parameters $x_s$
and $z$ are such that the surface $\Sigma$ is limited by ${\hat
E}=1$ plane and the surface ${\hat Q}=({\hat L}-a {\hat E})^2$
as is shown above. In figure (\ref{estrella}) we depict the
intersections of these surfaces, including slices obtained cutting
$\Sigma$ by planes of $z=constant$, as we explain below, for a particular values of $a$.

In this figure we can see how $\Sigma$ is in fact a surface folded
by the line from point $A$ to point $C_3$ made of singular points. The
equation of this line has been found in (\ref{ze}), (\ref{xize}) and
(\ref{Qze}). Each point represents the spherical orbit with the
maximum value of ${\hat Q}$ and the minimum of ${\hat L}$ for a
fixed value of $z$ and $a$.

Unstable spherical orbits are represented by the points of the lower
pseudo-trapezoid face with curved sides (that we represent by
a widehat over the extrema of the piece of curve). These sides are:
\begin{itemize}
  \item $\widehat{A,E_1}$ and $\widehat{C_3,E_{32}}$, unstable circular orbits on the equatorial plane,
  \item $\widehat{A,C_3}$, singular points of $\Sigma$,
  \item $\widehat{E_1,E_{32}}$, unstable spherical orbits with ${\hat E}=1$.
\end{itemize}

Stable spherical orbits form a partially unlimited face with
sides:
\begin{itemize}
  \item $\widehat{A,C_3}$, singular points of $\Sigma$,
  \item $\widehat{A,E_2}$ and $\widehat{C_3,E_{31}}$, stable circular orbits on the equatorial plane.
\end{itemize}

The apexes of these partially non-limited volume are:
\begin{itemize}
  \item $A$ the circular orbit in the equatorial plane with smallest value of ${\hat E}$ for a fixed value of $a$.
  \item $E_1$ and $E_{32}$ circular unstable orbits in the  equatorial plane with ${\hat E}=1$,
  \item $C_3$ circular orbit in the equatorial plane with smallest value of ${\hat E}$ and ${\hat L}<0$.
  \item $E_2$ and $E_{31}$ limiting points representing stable circular orbits in the equatorial plane when ${\hat E}\rightarrow 1$
\end{itemize}

We complement our knowledge of $\Sigma$ cutting slices of ${\hat
E}=constant (\Leftrightarrow z=constant)$. Then, the spherical
orbits could be represented in each slice in a parametric way by
(\ref{QLE}) where $z$ takes a constant value. We include the
intersection of each slice with the surface of the equatorial plane.

The complexity of the behavior of these parametric functions forces
the study of their particular shape for the cases studied in table
1. Relations listed in the previous chapter allow us to
better understand those particular shapes. \\

\begin{itemize}
\item {\bf Case 0}: $z=1$. $\widehat{E_1, E_{32}}$ unstable
spherical orbits with ${\hat E}=1$
\item {\bf Case I}: $1<z<z_b$.$\widehat{D_1,D_{32}}$ unstable
spherical orbits, $\widehat{D_2,D_{31}}$ stable spherical orbits,
\item{\bf Case II}: $z=z_b$.$\widehat{C_1,C_3}$ unstable spherical
orbits, $\widehat{C_2,C_{3}}$ stable spherical orbits ${\hat
E}=z_b^{-1/2}$,
\item {\bf Case III}: $z_b<z=z_c<z_a$ $\widehat{B_1,B_3}$ unstable
spherical orbits, $\widehat{B_2,B_{3}}$ stable spherical orbits
${\hat E}=z_c^{-1/2}$,
\item {\bf Case IV} : $z = z_a$ Last stable spherical orbit.
\end{itemize}

\begin{figure}
\begin{center}
\includegraphics[width=0.5\textwidth]{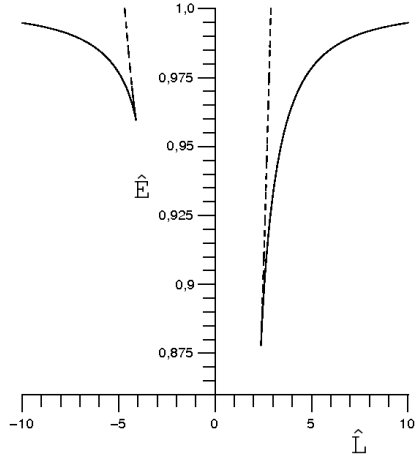}
\end{center}
\caption{\label{L-E}${\hat L} - {\hat E}$ plot of circular orbits on
the equatorial plane, ${\hat Q}=({\hat L}-a)^2$. }
\end{figure}

In order to complement the previous plots, we add the projection on
the ${\hat E}$,${\hat L}$ plane of the spherical orbits on the
equatorial plane. We represent these orbits on a parametric plot
using  ($\ref{xieq}$) into ($\ref{QLE}$). See figure
(\ref{L-E}) plotted for $a=0.8$, that are the projections of
$\widehat{A,E_1}$, $\widehat{A,E_2}$, $\widehat{C_3, E_{31}}$ and
$\widehat{C_3,E_{32}}$.

As $a$ decreases both peaks of this ${\hat L}-{\hat E}$ plot become
more and more symmetric until we arrive to $a=0$, where they become
completely symmetric (as could be expected from the spherical
symmetry of Schwarzschild black holes)

{\bf C} As we have seen in section D of the previous chapter,
in the general case I, the function $\mathcal{Q}_{s2}$ has
four values of $x_s$, say $x_i$ $i\in\{1,2,3,4\}$, such that
$\mathcal{Q}_{s2}=(a-\xi_{s2})^2$, for a fixed value of $a$ and $z$.
The plots have been obtained in all cases increasing
the value of the $x_s$ parameter, first from $x_1$ to $x_2$ (in figure \ref{estrella} from $D_1$ to $D_{32}$) for the
unstable branch, and from $x_3$ to $x_4$ for the stable one (from $D_{31}$ to $D_2$).
In the
other cases the end of the unstable and the start of the stable
branches is $x_{e}$ which traces the singular points of the global
surface.

We can now show the relative values
of $\xi_{s2}(x_i,z,a)$, that is

\begin{center}
$
  \begin{array}{cccc}
   \ & x_{i} & \xi_{s2}(x_i,z,a)\equiv \xi_i \\
   I&x_1 < x_2 < x_3 < x_4 & \xi_3< \xi_2<0<\xi_1<\xi_4 \\
  II&x_1 < x_2 = x_3 < x_4 & \xi_3= \xi_2<0<\xi_1<\xi_4 \\
  III&x_1 < x_4 & 0<\xi_1<\xi_4 \\
  IV&x_1 = x_4 & 0<\xi_1=\xi_4.
  \end{array}
$
\end{center}

Moreover, since $\xi_{s2}$ is a continuous function, it has a minimum in $x_{e}$
and takes positive values in the extrema of $\mathcal{D}$, $x_+$ and
$x_{s_{max}}$, from the above table, we can infer that $\xi_{s2}=0$
in two points $x_l$ ($\xi'_{s2}(x_l,z,a)<0$) and $x_m$
($\xi'_{s2}(x_m,z,a)>0$ and $x_l<x_m$) if $1<z\leq z_c$. Thus:
i) $x_1<x_l<x_2 < x_3<x_m<x_4$ in the case I and II, ii) if
$z_b<z<z_c$, then $x_{e} < x_l < x_1$ and $x_{e}< x_m< x_4$, iii) if
$z=z_c$ then $x_{e} = x_l=x_m$.

{\bf D} {\it Continuously moving  in a geometrical locus of
spherical orbits}. Let us choose $\textit{P}$, a point
of $\Sigma$. The holonomic base of the tangent space in $\textit{P}$
associated to curves on $\Sigma$ of $x_s=constant$ and $z=constant$
is
\begin{eqnarray}
{\bf X}_{z} &\equiv&\frac{\partial }{\partial z}= {\hat Q}_z \frac{\partial}{\partial {\hat Q}}+ {\hat
L}_z \frac{\partial}{\partial {\hat L}}+ {\hat E}_z
\frac{\partial}{\partial{\hat E}}\\
{\bf X}_{x_s} &\equiv&\frac{\partial }{\partial x_s}= {\hat Q}'\frac{\partial}{\partial {\hat Q}}+ {\hat
L}'\frac{\partial}{\partial {\hat L}},
\end{eqnarray}
where ${\hat Q}'$ and ${{\hat Q}}_z$ means partial differentiation
of ${\hat Q}_{s2}$ with respect $x_s$ and $z$ respectively and
$\frac{\partial}{\partial {\hat Q}}, \frac{\partial}{\partial
{\hat L}},\frac{\partial}{\partial {\hat E}}$ is the holonomic base on
$\textit{P}$ of  ${\hat Q}, {\hat L}, {\hat E}$ space.

As a consequence of gravitational radiation, in the adiabatic
approximation, a test particle can follow a trajectory that can be
imagined as a chain of orbits in such a way that the particle remains
several periods on each orbit before changing to another orbit. We
define ${\bf t}$ as an average over several orbital periods.
According to this, the constants of motion (${\hat Q}, {\hat L},
{\hat E}$) associated to different orbits vary according to a law
that evolves $ d{\hat Q}/d{\bf t},d{\hat L}/d{\bf t},d{\hat E}/d
{\bf t}$ quantities. In fact, the solution of the first order
ordinary equations is a curve in our formal space such that
\begin{equation}\label{}
{\bf V} = \frac{d {\hat Q}}{d{\bf t}} \frac{\partial}{\partial {\hat Q}}+
\frac{d {\hat L}}{d{\bf t}} \frac{\partial}{\partial {\hat L}}+ \frac{d {\hat E}}{d{\bf t}}
\frac{\partial}{\partial{\hat E}},
\end{equation}
where ${\bf V}\equiv d/d {\bf t}$, is its tangent vector ($\dot{\ }
\equiv d /d {\bf t}$). The necessary and sufficient
condition to ensure us that a particle in spherical motion remains
in a spherical orbit under the influence of radiation reaction is
\begin{eqnarray}\label{}
&\ &\epsilon_{ijk} V^i {X_{x_s}}^j {X_{z}}^k =0 \Leftrightarrow\\
{\hat L}' \frac{d {\hat Q}}{d{\bf t}}&=&\frac{{\hat Q}_z{\hat L}'-{\hat Q}'{\hat
L}_z}{{\hat E}_z} \frac{d {\hat E}}{d{\bf t}}+ {\hat Q}'\frac{d {\hat L}}{d{\bf t}},\label{rad}
\end{eqnarray}
where $\epsilon_{ijk}$ is the Levi-Civita tensor.
Hence, equation (\ref{rad}) becomes \cite{Kennefick}, (see also
equation 3.25 of \cite{Sago})
\begin{equation}\label{radsago}
\frac{d {\hat Q}}{d {\bf t}}=2\frac{2 x_s
+\sqrt{-2x_s(-z-2x_s+2zx_s)}}{(2x_s-1)\sqrt{z}}[(a^2+4x_s^2){\frac 
{ d{\hat E}}{d {\bf t}}}-a{\frac {d {\hat L}}{d {\bf t}}}].
\end{equation}

If the motion lies entirely on the equatorial plane then equation
(\ref{rad}), taking into account (\ref{Qequ}), becomes
\begin{equation}
\frac{d {\hat E}}{d{\bf t}} {\hat L}_z= \frac{d {\hat L}}{d{\bf t}} {\hat E}_z.
\end{equation}
The result is
\begin{equation}
\frac{d {\hat L}}{d{\bf t}}= -\frac{d {\hat E}}{d{\bf t}}\frac{1}{(2x-1)a}(4x^2+a^2
+\frac{2^{3/2} x^2 \Delta}{\sqrt{-x(2zx-2x-z)}}).
\end{equation}

Using basic instruments of classical differential geometry we can find the necessary and sufficient condition for a stable spherical geodesic be able to achieve a non null eccentricity orbit following ${\bf V}$ direction, that is, $\epsilon_{ijk} V^i {X_{x_s}}^j {X_{z}}^k <0$. This result applies if the start point is an unstable spherical orbit and we want to achieve a non null eccentricity orbit in ${\bf V}$ direction. In general this construction allow us if whether or not a specific curve, with ${\bf V}$ as a tangent vector in each point, is inside this partially unlimited volume of orbits in Kerr space time, and therefore if this is able to describe the path of a radiating particle in the adiabatical approximation.

\section{Non-null eccentricity Orbits in the ${\hat Q}-{\hat L}-{\hat E}$ space}

In the previous section we have constructed a space where we have
placed the spherical orbits, now we are going to study all orbits
and we will see that spherical ones delimitate where the rest of the
orbits are.

{\bf A} {\it The U potential}.
In order to analyze the non null eccentricity orbits in relation to the spherical ones we could use the potential technique factorizing $R$ as follows, \begin{equation}\label{hatU}
R=T({\hat U}-{\hat Q}),
\end{equation}
where $T$ and ${\hat U}$ are defined below. The complexity of ${\hat U}$ advise us to use $\eta$ to factorize
$R$. It will be easy to translate the results
in terms of ${\hat Q}$ and ${\hat L}$ as we show below.

According to (\ref{R2}) we have
\begin{eqnarray}
\label{R_w_U} R &=& \frac{\Delta}{4}[U-\eta],\label{Ug} \\
U&=& \frac{4 x^2}{x-1}[z-(z-1) x]- \frac{4
x(x-1)}{\Delta}(\xi+\frac{a}{x-1})^2=\nonumber\\
&=& -\frac{4x\left\{ 4(z-1) {x}^{3}-4z{x}^{2}+
\left[{\xi}^{2}+(z-1){a}^{2} \right]
x-({\xi}-{a})^{2}\right\}}{4\,{x}^{2}-4\,x+{a}^{2}}  \label{U}.
\end{eqnarray}
Therefore, considering that $T({\hat U}-{\hat Q})=\frac{\Delta}{4}(U-\eta)$, we obtain
\begin{eqnarray}
{\hat U} &=& z U-z(\xi-a)^2, \label{hatU-U} \\
T&=&\frac{\Delta}{4z},\nonumber\\
{\hat Q}&=& z^{-1}[\eta+(\xi-a)^2] \label{hatQ}.
\end{eqnarray}

Orbits exist only if $R\geq 0$. This implies
\begin{equation}\label{RU}
U \geq \eta.
\end{equation}

For fixed values of $\xi$ and $z$ parameters, and $a$, $U$ depends
on $x$.
\begin{itemize}
\item $x\rightarrow x_+$ then $U\rightarrow +\infty$
\item $x\rightarrow\infty$ then $U \rightarrow -\infty$
\item $U$ is continuous in the domain: $0<a\leq1$, $1<z$, $x_+<x<\infty$
\item Using Descartes's rule on the third-degree polynomial part (see(\ref{U})), one can
see that, for the values of $a$, $z$ and $\xi$ that we are
interested in, there are always 3 changes of sign. Thus there's the
possibility of having at most 3 real roots, and we know that there
will be at least 1 root, because this function, being continuous,
goes from $\infty$ to $-\infty$ inside that region.
\end{itemize}

Extrema of $U$ can't be found analytically. As it has been
explained, the conditions for spherical orbits are $R=0$ and $R'=0$.
In $U$ potential terms, they are equivalent to:

\begin{eqnarray}
\eta&=&U(x,a,\xi,z)\\
\label{Up}U'(x,a,\xi,z)&=&0
\end{eqnarray}
The set of values of $\xi$, $\eta$, $z$, $x$, $a$ which correspond
to spherical orbits are related through the functions $\xi_{s1,2}$
and $\eta_{s1,2}$. We are going to study the extrema of $U$ through
these functions proceeding as follows: First, one value of $a_0$ and
$z_0$ is chosen. Then, to completely define the $U$ potential, we
have to choose a particular value of $\xi$, say $\xi_0$.

\begin{figure}
\centering
\includegraphics[width=0.65\textwidth]{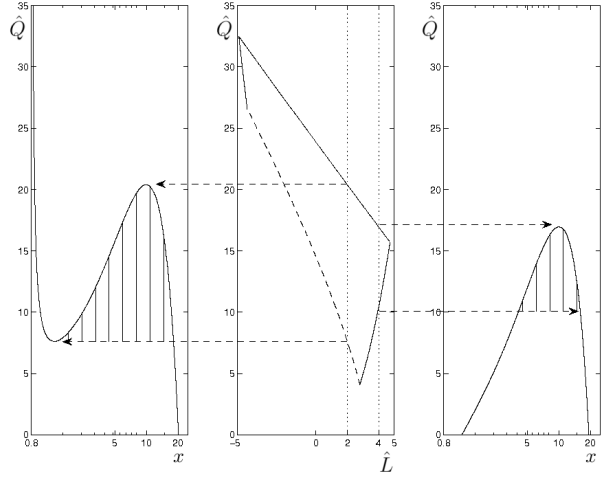}
\caption{\label{xi-eta-to-U}Relationship between ${\hat Q}-{\hat L}$
plot and ${\hat U}$ potential in two scenarios: at left ${\hat
L}_0=2$ intersects the stable and unstable curves and so we have
both extrema of ${\hat U}$ above the minimum value of ${\hat Q}$,
i.e $1.48$ $(\Leftrightarrow \eta=0)$. At right ${\hat L}_0=4$ only
intersects the stable orbits curve and then we only have the maximum
of ${\hat Q}$ above the minimum allowed value of ${\hat Q}$,
$10.36$. Regions where orbits exist are filled with vertical lines.}
\end{figure}

The two functions $\xi_{s1,2}$ coincide at the extrema of the
interval of $x_+ \leq x \leq x_{s_{max}}$. They are continuous
functions and each one has one extremum in the interval, a minimum
for $\xi_{s2}$ and a maximum for $\xi_{s1}$. So they form a oval
closed shape. Hence,  we can find different scenarios depending on
$\xi_0$:
\begin{itemize}
\item $\xi_0$ is such that $\xi_0 >\xi_{s1}$ or $\xi_0<\xi_{s2}$. Thus $U$ doesn't have any
extremum.
\item $\xi_0$ is tangent to the oval shape at $\xi_{s1}$ maximum or
$\xi_{s2}$ minimum. In this case $U$ potential has an inflection
point.
\item $\xi_0$ intersects twice the shape. This implies that two extrema exist in D.
In this case, intersections with $\xi_{s1}$ curves implies that the corresponding extrema are below the horizontal axis since $\eta_{s1}<0$.
But for intersections with
$\xi_{s2}$ the extremum will be over or below the horizontal axis
depending on the sign of $\eta_{s2}$.
\end{itemize}

As we have seen, $U$ is a continuous function that takes the values
$+\infty$ and $-\infty$, at $x_+$ and $x\rightarrow \infty$,
respectively. Then if it has two extrema, it must first have  a
minimum, and after a maximum, with the maximum being above the
minimum.

In figure (\ref{xi-eta-to-U}) we can see the more general example where the two possible scenarios are plotted
in terms of ${\hat Q}$ and ${\hat L}$ by using (\ref{hatU-U}), and (\ref{hatQ}) (the form of ${\hat U}$ is like $U$ except a shift and a rescaling).
We have complemented the information comparing these two plots with
the corresponding ${\hat Q}$ versus ${\hat L}$, that corresponds to the Case I (slice $ D_1,D_2,D_{31},D_{32}$ of figure (\ref{estrella})
of the previous section), establishing graphical analogies.

Now that we know the shape of ${\hat U}$ we can find where the
non-spherical orbits are. An orbit must exist between two turning
points, if we choose a value ${\hat Q}_0$  we see that it can
intersect up to three times the ${\hat U}$ potential defining three
turning points $x_t \leq x_a \leq x_p$. We can find orbits only when
${\hat Q}_0$ value is between both ${\hat U}$ extrema (considering
${\hat Q}_0 \geq ({\hat L}_0-a{\hat E}_0)^2$ , where ${\hat L}_0$
and ${\hat E}_0$ are fixed values defining ${\hat U}$) and $x_a \leq
x \leq x_p$, such that the apoaster is in $x_a$, and the periaster
is in $x_p$. This situation corresponds to the region between the
two curves at the ${\hat Q}$ versus ${\hat L}$ plots and the curves
$\eta=0\Leftrightarrow {\hat Q}=({\hat L}-a{\hat E})^2$. Geodesics
falling into the hole exist in $x_+\leq x \leq x_t$ region
(associated geodesics \cite{Chandra}),  assuming the same values of
the physical parameters ${\hat Q}_0$, ${\hat L}_0$ and ${\hat E}_0$
of the associated orbits.

Figure (\ref{xi-eta-to-U}) tell us that, starting from a value of
${\hat Q}_0$ corresponding to an orbit, if we increase continuously
this value we can achieve at most the value ${\hat U}_{max}$ in
order to have physical orbits. But if we decrease this value in the
left case of (\ref{xi-eta-to-U}) below ${\hat U}_{min}$ we will get
open geodesics that will be swallowed by the hole, but at the right
case of (\ref{xi-eta-to-U}), if ${\hat Q}_0<({\hat L}-a{\hat E})^2$
no physical orbits exist. Therefore, the surfaces representing
stable spherical orbits and  the ${\hat Q}_0=({\hat L}-a{\hat E})^2$
surface in figure (\ref{estrella}) are impenetrable, i.e. can be
achieved but not crossed following a continuous change of the
physical parameters from an interior point of the volume of orbits,
while the surface of unstable spherical orbits can bidirectionally
be crossed.

{\bf B} {\it The region where orbits exist}.

If we focus on  $1<x$ region we can obtain
interesting results due to the fact that, in that region,
\begin{equation}\label{}
0\leq \eta \leq U(\xi,z,x,a) \leq h(x,z) \label{etag}
\end{equation}
since
\begin{eqnarray}\label{}
U&=&h-\frac{4 x(x-1)}{\Delta}(\xi+\frac{a}{x-1})^2 \\
h&=& \frac{4 x^2}{x-1} [z-(z-1)x].
\end{eqnarray}

If $x\rightarrow1+$ then
$h\rightarrow+\infty$, if $x\rightarrow+\infty$ then
$h\rightarrow-\infty$, besides  if $1<z\leq 9/8 $ has a minimum in
$x_{e1}$ and a maximum in $x_{e2}$, $x_{e1}\leq x_{e2}$ and
$h(x_{e1},z)\leq h(x_{e2},z)>0$, and if $9/8<z$, thus $h(x,z)$ is a
continuously decreasing function. In addition to this $h(x,z)\geq 0
\Leftrightarrow 1< x \leq x_{limit}$, where
\begin{equation}\label{xlimit}
x_{limit}\equiv z/(z-1).
\end{equation}
Considering that $h(x,z)<0$ in the $x_+\leq x\leq 1$ region, we can
assert that the allowed values of $x$ coordinate of orbits are
bounded by $x_+<x \leq x_{limit}(z)$. The upper limit only depends
on mass energy ratio.

{\bf C} {\it Orbits co rotating with the hole}

According to figure $\ref{estrella}$, physical orbits with  $z\geq
z_c$ are characterized by ${\hat L}\geq 0$ ($\Leftrightarrow \xi\geq
0$). In fact, there exists a set of orbits with the same value of
${\hat L}$. These orbits are placed in this figure in the slice of
${\hat L}=constant \geq 0$, including a number of stable and
unstable orbits. Therefore Theorem IV must be extended beyond
spherical orbits, concluding that all orbits with $z\geq z_c$ co
rotate with the hole. That is, following (\ref{de3p}) we can check
that $d\phi/d\tau \geq 0$ for all orbits with $z\geq z_c$.
\begin{equation}
\frac{z^{\frac{1}{2}} {\bar \rho}^2}{2} \frac{d \phi}{d {\bar \tau}}=
(\frac{1}{1-\mu^2}-\frac{a^2}{\Delta})\xi +\frac{4 a x}{\Delta} \geq  (1-\frac{a^2}{\Delta})\xi +\frac{4 a x}{\Delta}.\nonumber
\end{equation}
But, we can always find a spherical orbit such that $\xi=\xi_{s2}$.
Thus
\begin{equation}
(1-\frac{a^2}{\Delta})\xi +\frac{4 a x}{\Delta}-\xi=(1-\frac{a^2}{\Delta})\xi_{s2} +\frac{4 a x}{\Delta}-\xi_{s2} = {\frac {a \left( 1+\sqrt {2}\sqrt {- \left( -2\,x+2\,zx-z \right) x}\right) }{2\,x-1}}> 0. \nonumber
\end{equation}
Therefore
\begin{equation}
\frac{z^{\frac{1}{2}} {\bar \rho}^2}{2} \frac{d \phi}{d {\bar \tau}}=(\frac{1}{1-\mu^2}-\frac{a^2}{\Delta}) +\frac{4 a x}{\Delta} > \xi.
\end{equation}

\section{Summary of main results}

The main goal of this paper has been to provide different tools in
order to have a very important knowledge of both (unstable and stable) spherical orbits, and non null
eccentricity orbits in the outer Kerr space time, studying how their physical parameters are related. This is due to the special
role played by Kerr orbits to explain astrophysical effects in
strong gravitational fields such as gravitational wave radiation in
the adiabatic approximation.

To achieve this, we have developed an analytical  study
of orbits out of the equatorial plane, recovering, as expected,
classical results of orbits lying entirely on the plane of symmetry
including the Schwarzschild results and the limit of Kerr metric
($a=1$). A numerical approach to obtain specifical results about
orbits or radiative models can be implemented based on our
analytical approach.

1) We have classified spherical orbits in 4 different classes
according to the different behavior of the physical parameters restricted to spherical orbits
${\hat Q}_{s2},{\hat L}_{s2}$ and ${\hat E}$.  The influence of $a$
in the results has been studied, as well as maximum values, domain
of existence and correlations of ${\hat Q}_{s2}$ with ${\hat
L}_{s2}$, showing that these two functions have a common extremum in
$e$ ( a minimum of ${\hat L}_{s2}$ and a maximum of ${\hat Q}_{s2}$)
that, at the same time, is the threshold that divides stable and
unstable spherical orbits (see Theorem II).

2)Moreover, using the properties of the extremum $e$ mentioned above, we have
found that, for $z>z_{c}(a)$ ($\Leftrightarrow {\hat L}_{s2}\geq
0$), all orbits (including the spherical ones) have a positive value
of ${\hat L}$ and therefore must co-rotate with the hole.

3)According to this classification we have constructed the physical parameters space
(${\hat Q},{\hat L},{\hat E} $) showing that the whole
set of spherical and non-null eccentricity orbits can be represented
there. We have been able to carry this out thanks to the fact that
 the Kerr metric has two killing vectors and one killing
2-tensor and the choice of Boyer-Linquist coordinates. In this space
the geometrical locus of spherical orbits is a 2-surface that looks
like two sheets (one for stable and the other for unstable spherical
orbits), partially matched through a common curve. We have proved
that this curve is the set of singular points of this 2-surface
characterized by the condition ${\hat Q}'_{s2}={\hat L}'_{s2}= R''=0
\Leftrightarrow x_s=x_{e}(z,a)$.

This surface can be constructed using slices of $z=constant$. Each
slice exhibits the limits of ${\hat Q}$ and ${\hat L}$ values,
showing that beyond $z_a (a)$ no physical orbits ($\eta\geq
0$) exist. It means that if $E < E_a=m \sqrt{z_a}$ neither spherical nor
non null eccentricity orbits exist. The larger value of $z_a$ is $3$
for $a=1$.

4) Studying the $U$ potential we have seen that non-null
eccentricity orbits are placed in this formal three-space between
the two sheets of stable and unstable spherical orbits and limited
by the surface representing the equatorial plane. Then, we obtained
a global representation of the whole set of orbits in the outer
region of the Kerr space-time. This representation allows us to see
whether or not a particle can achieve  one specific orbit from
another one with a continuous change of its parameter values.

5) Using a property of $U$ potential, we have obtained $x_{limit}$
(see (\ref{xlimit})), that is, a maximum  value for the  $x$
coordinate of an orbit (spherical or not) depending on its
energy-mass relation, ${\hat E}$.

6) We have obtained the necessary and sufficient condition that the variations of
the ${\hat Q}, {\hat L}$ and  ${\hat E}$
parameters must hold so that a spherical orbit remains spherical.
Using this condition, we recover the result obtained in
\cite{Apostolatos} in the Schwarzschild limit and small
perturbations of a stable spherical orbit, and in \cite{Kennefick} in
Kerr space-time.

7)Variations of the parameters due to gravitational radiation in the
adiabatical approximation can be represented in this formal space as a
curve going form one geodesic to another. Therefore we can check
whether or not this path is possible according to the fact that the
upper sheet of $\Sigma$ (that of stable spherical orbits) is
impenetrable while the lower one is not, as has been proved in
section 6. This representation, as a formalization of analytical
results, can be an important tool to check whether a transition
from one orbit to another is valid or not.

In particular the necessary condition to go from one stable
spherical geodesic to an unstable one (or vice versa) following a way that only
include spherical orbits is crossing the
singular line of $\Sigma$.

\section{Appendix I}

In order to prove that $\eta_{s1}<0$ in $\mathcal{D}$, we can do the next
changes of variable and parameter: a) $z=2 x_{s_{max}}/(2
x_{s_{max}}-1)$, b) $x_s=t_s+ x_+$ and c) $x_{s_{max}}=t_{smax}+x_+$
(then $0\leq t \leq t_{smax}$) where $x_{s_{max}}$ is defined in
(\ref{xsmax}) and $x_+$ is the outer horizon. After applying this
changes, we obtain
\begin{equation}
\eta_{s1} = -\frac{4 x_s^2}{(2x_s-1)^2 a^2}[J(x_s,a,z)+2 \Delta
\sqrt{2x_s(2(1-z)x_s+z)}],\nonumber
\end{equation}
where
\begin{eqnarray}
J(x_s,a,z)&\equiv& j(t_s,a,t_{smax})=\nonumber\\
&+&\frac{8}{2 t_{smax}+\sqrt{1-a^2}} [ 2(t_{smax}-t_s) +4] t_s^3 +\nonumber\\
&+& 4[6\sqrt{1-a^2}(t_s-t_{smax}) +(3(\sqrt{1-a^2}-2))]t_s^2\nonumber\\
&+& 4[(1-a^2)+ 4
(1-a^2)t_{smax}+\sqrt{1-a^2}]t_s+\nonumber\\
&+&(1-a^2)[4t_{smax}+4t_{smax}\sqrt{1-a^2}+2\sqrt{1-a^2}+(2-a^2)]\nonumber\\.
\end{eqnarray}
This implies that $\eta_{s1}<0$ in the intervals of interest,
q.e.d..

\section{Appendix II}

After some manipulations we can see that the system of equations
$\xi'_{s2}=\mathcal{Q}'_{s2}=0$ is equivalent to
\begin{equation}\label{aG}
 0=a^2 + G(x_s,z),
\end{equation}
where
\begin{eqnarray}\label{a2xe3}
G(x_s,z)&=& -\frac{4x_s}{z^2}\big\{ 8(z-1)(z-2)x_s^2+ 3z(4-3z)x_s +3z^2+ \nonumber\\
&+& 4[2(z-1)x_s-z]\sqrt{2x_s[2(1-z)x_s + z]} \big\}.
\end{eqnarray}
In fact, functions $\xi'_{s2}$ and $\mathcal{Q}'_{s2}$ can be
written
\begin{eqnarray}\label{}
\xi_{s2}'(x_s,a,z)&=&\frac{H(x_s,z)}{a}[a^2 + G(x_s,z)],\\
\mathcal{Q}_{s2}'(x_s,a,z)&=& \frac{F(x_s,z)}{a^2}[ a^2 + G(x_s,z)],
\end{eqnarray}
where $F(x_s,z)\neq 0$  and $H(x_s,z)\neq 0$.

We define functions $S(x,a,z)$ according to
\begin{eqnarray}
  S(x,a,z)&=& G(x,z)+a^2= \\
  &=& a^2-\frac{4x}{z^2}[8(z-1)(z-2)x^2 + 3z(4-3z)x+3z^2]-, \nonumber\\
  &-&\frac{16x}{z^2}[2(z-1)x-z]\sqrt{-2x[2(z-1)x-z]}.\label{Sx}
\end{eqnarray}

We implement the change of variable
\begin{equation}
x\rightarrow y=\frac{x}{x_{s_{max}}}-1/2, \label{yx}
\end{equation}
where $x_{s_{max}}$ has been defined in (\ref{xsmax}).

This change implies that $x=0 \leftrightarrow y=-1/2$ and
$x=x_{s_{max}}\leftrightarrow y=+1/2$. At the same time $y=0
\Leftrightarrow x= x_{s_{max}}/2$. In addition to this
\begin{equation}\label{}
x_+ \leftrightarrow y_+=
\frac{1}{2}-\frac{1}{z}+(1-\frac{1}{z})\sqrt{1-a^2}.
\end{equation}
Now, the position of the outer horizon $y_+$ depends not only on $a$
but also on $z$. Remember that the domain $\mathcal{D}$  exists iff. condition
$x_+ \leq x_{s_{max}},\ \  \Leftrightarrow 1-(z-1)\sqrt{1-a^2}\geq
0$ holds.

The dependence of function $S$ on $y$ is
\begin{eqnarray}
S(y,z,a)&=&a^2-\frac{z}{(z-1)^2}[4 (z-2) y^3 - 3z y^2 + \frac{5}{4}
z-1]+\\
&+&\frac{z}{(z-1)^{3/2}} (1-4 y^2)^{3/2}.
\end{eqnarray}

The values of $S$ in the following points are
\begin{eqnarray}
  S(y=-\frac{1}{2}) &=& a^2\geq 0, \label{Sm12}\\
  S(y=+\frac{1}{2}) &=& a^2-1+\frac{1}{(z-1)^2}=\frac{(1-(z-1)\sqrt{1-a^2})^2}{2(z-1)\sqrt{1-a^2}} \geq 0,\label{SM12}\\
  S(y=0)&=&
  -a^2+\frac{z[5\,(z-1)-4\,(z-1)^{1/2}+1)]}{4(z-1)^2} <0,\\
  S(y_+)&=&-\frac{4}{z^2}\{(1-(z-1)\sqrt{1-a^2})(1+\sqrt{1-a^2})\times\nonumber\\
  &\times&[\sqrt{1-(z-1)\sqrt{1-a^2}}-\sqrt{1-a^2}]^2\}< 0 .\label{S0}
\end{eqnarray}
As a direct consequence we see that $S(\pm 1/2)\geq 0$, $S(0)$ is a
negative function of $z$, ($1<z$), such that $S(0)<a^2-1$, and
$S(y_+)\geq 0$.

Moreover,
\begin{equation}\label{}
S'(y)=\frac{6 z y}{(z-1)^2}[2(z-2)y-z+2(z-1)^{1/2} (1-4y^2)^{1/2}].
\end{equation}
The equation $S'(y)=0$ has two solutions in
\begin{eqnarray}
y_{1}&=&0, \nonumber \\
y_{2}&=&\frac{1}{2}-\frac{1}{z}, \ \ \  S(y_2)=a^2 -1.
\end{eqnarray}
It is crucial to see that
\begin{equation}\label{y+y2}
 y_+=y_{2} +(1-\frac{1}{z})(1-a^2)^{1/2}\ \ \  \Rightarrow y_+\geq
 y_{2},
\end{equation}
and $S(0)=S(y_1)<S(y_2)\leq 0$.

Considering these previous results, and taking into account the
continuity of $S$, we can assert that this curve has at least
one root ($S=0$) in the interval $-1/2<y\leq 0$ and at least another
one in $0\leq y\leq+1/2$. This is the proof of the existence of at
least two solutions of the equation (\ref{aG}). But we don't know if
there exist solutions in the interval $y_+\leq y \leq 1/2$, and if
there exist only one.

The "mean-value theorem" states that if there exist two points
$y_a, y_b$, $y_a< y_b$, such that $S(y_{a,b})=0$, then there is, at
least, a point $y_0$, $y_a< y_0 < y_b$, such that $S'(y_0)=0$. As we
have at least two roots of the equation $S(y,z,a)=0$,  then, if we
find $y_0$, we can assert that one root must be a value greater than
$y_0$ (and another one less than $y_0$). Therefore, if $y_+\leq y_0$
or $y_+\leq y_1$ then we will prove that one single cut point will
be outside the hole.

We assert that there is a solution beyond the outer horizon iff.
$y_+\leq y_{01}=0$, that is equivalent to
\begin{equation}\label{}
z\leq z_{limit}(a)\equiv \frac{1+ (1-a^2)^{1/2}}{\frac{1}{2}+
(1-a^2)^{1/2}}.
\end{equation}
since, according to (\ref{y+y2}) $y_{02}\leq y_+$. Nevertheless,
there exist spherical orbits with $z> z_{limit}$, so we
don't know if in this limiting cases the root of $S$ could be inside
the outer horizon. But this is not possible since $S(y_+)\leq
0$ and then the root must be outside the horizon (the equality meets
in the limit when $y_+=1/2$). \\

{\bf Acknowledgements} \\
We wish to thanks our colleagues Dr. J. Llosa and Dr. C.F. Sopuerta for their help and useful comments.

\end{document}